\documentclass[superscriptaddress,twocolumn,amsmath,amssymb,longbibliography]{revtex4-1} %floatfix
\usepackage{graphicx}
\usepackage{bm}
\usepackage{verbatim}
\usepackage{dcolumn}
\usepackage{color,etoolbox}
\usepackage{SIunits}
\usepackage{booktabs}
\usepackage{textcomp}
\usepackage{multirow}
\usepackage[colorlinks, linkcolor=blue, anchorcolor=blue, citecolor=red, urlcolor=blue]{hyperref}

\newcommand{\yso}{Y$_2$SiO$_5$}
\newcommand{\euyso}{$^{151}$Eu$^{3+}$:Y$_2$SiO$_5$}

\newcommand{\mm}{mm~}
\newcommand{\um}{$\upmu$m}

\newcommand{\us}{$\upmu$s}
\newcommand{\nth}{$\mathrm{n}^{th}$}

\newcommand{\ket}[1]{\vert#1\rangle}

\newcommand{\e}[0]{\mathrm{e}}

\providecommand{\enquote}[1]{#1}
\renewcommand{\enquote}[1]{#1}

\begin{document}
\title{Random access and high dimensional integrated quantum memory}

\author{Zhong-Wen Ou}
\altaffiliation[]{These authors contributed equally to this work.}
% \author{Tian-Xiang Zhu\footnote[2]{These authors contributed equally to this work.}}
\author{Tian-Xiang Zhu}
\altaffiliation[]{These authors contributed equally to this work.}
\affiliation{Laboratory of Quantum Information, University of Science and Technology of China, Hefei 230026, China}
\affiliation{Anhui Province Key Laboratory of Quantum Network, University of Science and Technology of China, Hefei 230026, China}
\affiliation{CAS Center For Excellence in Quantum Information and Quantum Physics, University of Science and Technology of China, Hefei 230026, China}
\author{Peng-Jun Liang}
\author{Xiao-Min Hu} 
\author{Zong-Quan Zhou}
\email[Correspondence: ]{zq\_zhou@ustc.edu.cn}
\author{Chuan-Feng Li}
\email[Correspondence: ]{cfli@ustc.edu.cn}
\author{Guang-Can Guo}
\affiliation{Laboratory of Quantum Information, University of Science and Technology of China, Hefei 230026, China}
\affiliation{Anhui Province Key Laboratory of Quantum Network, University of Science and Technology of China, Hefei 230026, China}
\affiliation{CAS Center For Excellence in Quantum Information and Quantum Physics, University of Science and Technology of China, Hefei 230026, China}
\affiliation{Hefei National Laboratory, University of Science and Technology of China, Hefei 230026, China}

% \date{\today}
\begin{abstract}
Random-access quantum memory (RAQM) is a fundamental building block for scalable quantum networks and photonic quantum computing. While existing realizations rely on gaseous physical systems, solid-state platforms—particularly integrated architectures—offer distinct practical advantages for scalable deployment. However, current integrated quantum memories are limited to single-channel operation, which precludes both RAQM and the manipulation of high-dimensional photonic states. Here, we demonstrate an 11-channel integrated quantum memory based on laser-written waveguide arrays in an \euyso~crystals. On-chip electrode arrays allow independent control of the read‑out time in each channel via Stark-shift-induced atomic interference. The device achieves random-access quantum storage of three time-bin qubits with a fidelity exceeding 99\%, and stores five-dimensional path-encoded quantum states with a fidelity above 96\%. This multichannel integrated quantum memory supports flexible functionality through its random-access operation and establishes a practical hardware platform for high-dimensional quantum networks implemented in integrated architectures.
\end{abstract}       

\maketitle

\section*{Introduction}
{
\color{black}
A random-access quantum memory (RAQM) \cite{jiang2019experimental,langenfeld2020network,zhang2024realization}, which stores multiple qubits in individually addressable channels and retrieves them on‑demand in any order, is essential for scalable quantum networks and photonic quantum computing. By enabling flexible entanglement connection between elementary links through qubit routing, RAQM can increase the rate of quantum repeaters while relaxing the demand on memory lifetimes and physical channels \cite{collins2007multiplexed,zhang2024realization}. It further permits the entanglement swapping between time-bin qubits with variable delays in a multiplexed quantum repeater based on time measurements \cite{halder2007entangling,zhu2026metropolitan}, offering capabilities beyond the linear speedup provided by static-access multiplexing \cite{seri2017quantum,seri2019quantum,tang2015storage,liu2021heralded,ortu2022storage,businger2022non}. 
Moreover, RAQM can substantially improve the efficiency of photonic quantum computing \cite{giovannetti2008quantum,briegel2009measurement,hosseini2009coherent,campbell2014configurable,liu2024nonlocal} by buffering asynchronously arriving photons to prepare large‑scale cluster states and multiphoton states \cite{nunn2013enhancing}. 

To date, RAQM has been only realized in gaseous physical system including cold atomic gases \cite{jiang2019experimental,zhang2024realization} and two single atoms coupled to an optical cavity \cite{langenfeld2020network}. For practical, scalable deployment, solid-state implementations—particularly those with integrated photonic structures \cite{saglamyurek2011broadband,zhong2017nanophotonic,zhou2023photonic}—are highly promising for achieving RAQM with high storage density. However, previous demonstrations of integrated quantum memories in rare-earth-ion-doped solids \cite{saglamyurek2011broadband,zhong2017nanophotonic,zhou2023photonic,seri2019quantum,liu2020demand,zhu2022demand,meng2026efficient,liu2025millisecond,liu2022demand} have been confined to single-channel operation, which precludes the implementation of RAQM.

A related challenge involves the storage of high-dimensional photonic quantum states (qudits). Qudits offer significant advantages over qubits, including higher channel capacity, enhanced noise robustness, improved security, and simplified quantum circuit complexity \cite{ecker2019overcoming,erhard2020advances,zheng2022entanglement,campbell2014enhanced,lanyon2009simplifying}. While integrated photonics \cite{li2021subwavelength,luo2023recent,ren2023photonic} already provides mature platforms for generating, manipulating, measuring, and distributing high‑dimensional entangled states \cite{wang2018multidimensional,zheng2023multichip}, an integrated memory capable of storing such states has remained missing, due to the difficulty of encoding high‑dimensional information in a single‑channel integrated structure.

In this work, we demonstrate the first solid-state RAQM using an 11-channel integrated quantum memory fabricated via laser-written waveguide arrays in an \euyso~crystal. These waveguide memory arrays are integrated with coplanar electrode arrays  that provide independent electrical control, allowing for the first demonstration of RAQM for time-bin-encoded qubits. Furthermore, by exploiting path encoding, we demonstrate reliable storage of high-dimensional photonic qudits in the same platform, establishing a key hardware foundation for high speed, high-dimensional quantum networks \cite{liu2021heralded,zhou2023photonic,erhard2020advances}.
}

\section*{Results}
\subsection*{Experimental setup}

\begin{figure*}[!t]
\centering
\includegraphics[width=\textwidth]{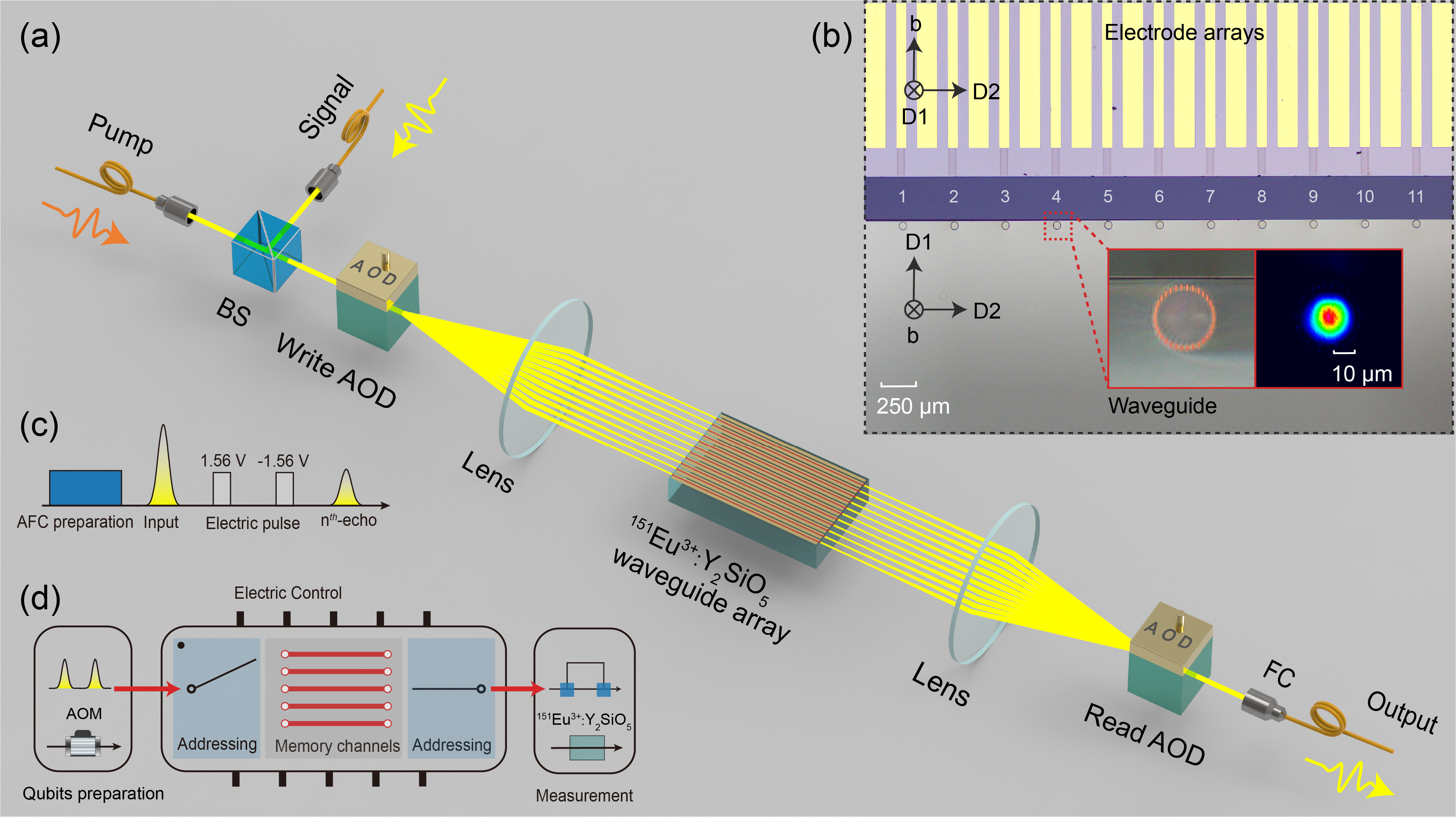}
\caption{Experimental setup. (a) Schematic of main optical setup. The pump beam and the signal beam are combined via a $90:10$ (R:T) beam splitter (BS). The combined beam is addressed to 11 memory channels based on optical waveguide array through the write acoustic optical deflector (AOD) and a lens. The optical polarization is aligned parallel to the crystal's $D1$ axis. The read AOD couples the output from 11 channels into a single-mode fiber via a fiber collimator (FC) for further single-photon analysis.  (b) Microscopic view of waveguide arrays and electrode arrays. The upper picture is a vertical view of the device, showing center electrodes with a width of $50$ \um~used to apply the electrical pulse, and neighboring electrodes with a width of $100$ \um~that are grounded. The lower image is a front view of the device, with small circles indicating the cross-section of each waveguide. Inset (red box): Enlarged view of the waveguide and the mode profile. (c) Time sequence for SMAFC storage. The \nth~AFC echo can be on-demand retrieved from the memory by applying a pair of electric field pulses. (d) The time-bin qubit storage architecture consists of a qubit generation unit (prepared via an acousto-optic modulator, AOM), a memmory unit with electric control, and a measurement unit (analyzed using another \euyso~crystal prepared with AFC).
}\label{fig1}
\end{figure*}

A microscopic image of the multichannel integrated quantum memory is depicted in Fig. \ref{fig1}(b). {\color{black} The substrate is a 0.07\% doped \euyso~crystal, a unique material that enables coherent light storage for 1 hour \cite{ma2021one} and single-photon-level storage for 1 minute \cite{lv2025minute}.} The substrate dimensions are $5 \times 4 \times 15 $ \mm along the crystal's $D1\times D2\times b$ axes. Eleven depressed cladding optical waveguides \cite{zhu2022demand} are fabricated along the $b$ aixs using a femtosecond laser micromachining (FLM) system. Electrode arrays are plated on the crystal's $D2\times b$ surface to individually apply electric fields to the Eu$^{3+}$ ions within the optical waveguides, allowing active control of the storage process. Further details about the device are provided in Sec. 1 of Supplemental Materials. We employ a pair of acoustic optical deflectors (AODs) to individually address the eleven memory channels (Fig. \ref{fig1}(a)). By applying controlled radio-frequency (RF) signals to the AODs, light can be programmatically directed to any single memory channel or their coherent superposition. 

The storage of photons is achieved using the Stark-modulated atomic frequency comb (SMAFC) protocol \cite{horvath2021noise,liu2020demand,zhu2022demand}, which absorbs single photons via an atomic frequency comb (AFC) \cite{de2008solid,afzelius2009multimode} and enables on-demand retrieval in discrete steps through active control of the atomic rephasing process by Stark-induced interference between two subgroups of atoms. Specifically, with an AFC prepared with a frequency periodicity of $\Delta$, an electric field pulse applied during the time interval $[0,1/\Delta]$ can suppress the standard AFC echo emission at $1/\Delta$ through complete destructive interference. A second electric field pulse with the opposite polarity applied during the time interval $[(n-1)/\Delta,n/\Delta]$ can then read out the photons at $t = n/\Delta$, corresponding to the \nth~AFC echo. The timing sequence for the SMAFC storage process is illustrated in Fig. \ref{fig1}(c). An electric pulse with a duration of $50$ ns and a voltage of $1.56$ V is sufficient for our integrated device, allowing for easy control using standard electronic devices. Fig. \ref{fig1}(d) depicts the time-bin qubit storage architecture, which allows multiple qubits to be stored and retrieved in any desired order.

\subsection*{Multichannel storage of single-photon-level inputs} 

\begin{figure*}[!t]
\includegraphics[width=1\textwidth]{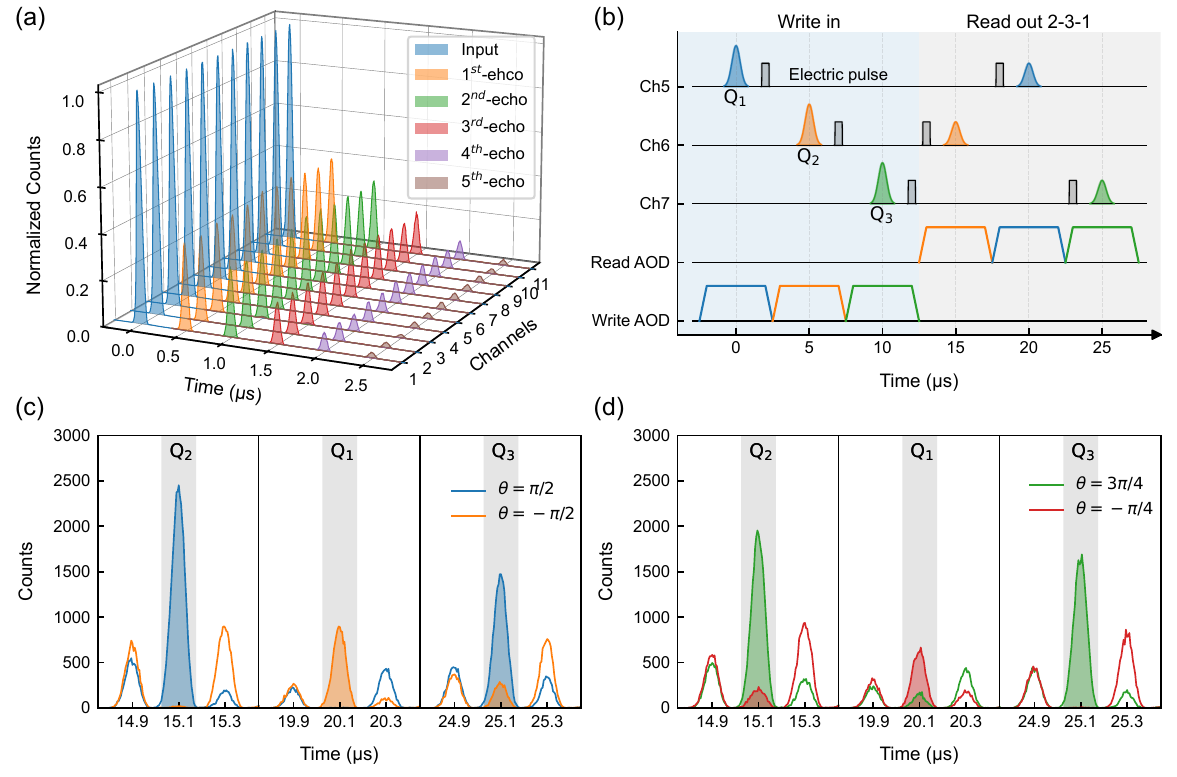}
\caption{Multiplexed and random access quantum storage of time-bin qubits. (a) Photon counting histograms for multiplexed storage of weak coherent inputs, demonstrating balanced efficiency for all 11 channels. On-demand retrieval of each channel is independently controlled via the corresponding electrodes, enabling precise and flexible operation. (b) The time sequence for random access quantum storage of three time-bin qubits is illustrated. The colors of the pulses represent distinct RF frequencies associated with the corresponding memory channels. The shown sequence exemplifies a specific retrieval order of 2-1-3, which can be reconfigured by adjusting the timing of the second electric pulses per channel, enabling arbitrary read-out orders.  (c,d) Interference measurements on the retrieved time-bin qubits are presented for the read-out order of 2-1-3. The results correspond to input qubits $Q_1: (\ket{e}-i\ket{l})/\sqrt{2},~ Q_2:(\ket{e}+i\ket{l})/\sqrt{2}$ and $Q_3:(\ket{e}+\mathrm{e}^{i3\pi/4}\ket{l})/\sqrt{2}$, with the readout basis defined as ($\ket{e}+{e}^{i\theta}\ket{l})/\sqrt{2}$. (c) The read-out phases are set to $\theta=\pi /2$ (blue solid line) and $\theta=-\pi/2$ (orange solid line). (d) The read-out phases are set to $\theta=3\pi/ 4$ (green solid line) and $\theta=-\pi /4$ (red solid line). The interference visibilities measured for $Q_2$, $Q_1$, and $Q_3$ within a 150-ns detection window are $98.5\pm0.1\%$, $99.3\pm0.1\%$, and $99.3\pm0.1\%$, respectively.} 
\label{fig2}
\end{figure*}

We first demonstrate multiplexed storage with eleven independent channels. The inputs are weak coherent pulses with a mean photon number per pulse of $\mu=0.14$. We set $\Delta=$ 2 MHz for all channels. Each channel allows on-demand retrieval controlled by corresponding electric pulses (Fig. \ref{fig2}(a)). The average storage efficiency for the 1$^{st}$/2$^{nd}$ AFC echo is $39.2\pm 0.2\% $/$31.3\pm 0.2\%$, The average storage efficiency for the 1$^{st}$/2$^{nd}$ AFC echo is $39.2\pm 0.2\% $/$31.3\pm 0.2\%$, surpassing the previous record of 27.8\% for integrated photonic memories \cite{liu2020demand}, with consistent storage performance across all 11 channels. Our device further features inter-channel crosstalk below -40 dB and electrical crosstalk-induced efficiency degradation below $1\%$ (see Sec. 2 in Supplemental Materials), making it well-suited for implementing more complex manipulations across many independent storage channels.

\subsection*{Random access storage of multiple time-bin qubits}

{\color{black}Static-access multiplexing has been extensively  studied in solid-state quantum memories \cite{seri2019quantum,tang2015storage,seri2017quantum,ortu2022storage,liu2021heralded,businger2022non,zhu2026metropolitan}, where sequences of photonic pulses are stored and retrieved in a fixed, first‑in‑first‑out order. In contrast, our multichannel memory—featuring fully independent control per channel—overcomes this limitation and enables RAQM, allowing any stored temporal mode to be read out on demand and in an arbitrary sequence.} We begin by initializing memory channels 5, 6, and 7 through the simultaneous preparation of AFC structures, with a comb spacing of $200$ kHz. Subsequently, the write AOD sequentially writes the time-bin qubits  $Q_1$, $Q_2$ and $Q_3$ into memory channels 5, 6, and 7, respectively. Finally, these three qubits can be read out in any desired order, as illustrated in Fig. \ref{fig2}(b), by controlling the timing of the electric pulses applied to each channel and the read AOD. Here we show an example to store distinct time-bin qubits $Q_1:(\ket{e}-i\ket{l})/\sqrt{2},~ Q_2:(\ket{e}+i\ket{l})/\sqrt{2}$ and $Q_3:(\ket{e}+\e^{i 3\pi/4}\ket{l})/\sqrt{2}$, where $\ket{e}$ and $\ket{l}$ represent the early and the late qubits, respectively, separated by an interval of $200$ ns. To analyze the superposition states, we employ another \euyso~crystal, prepared with an AFC structure, to serve as an unbalanced Mach–Zehnder interferometer \cite{kutluer2019time}. The AFC storage time is set to 200 ns to match with the time-bin interval and the interference phase is controlled by the AFC detuning (see Sec. 2 in Supplemental Materials for details). The fidelity of superposition states is defined as $F=(V+1)/2$, while $V$ is the interference visibility \cite{gundougan2015solid,liu2020demand}. Example interference results for $Q_1,Q_2$ and $Q_3$ are shown in Fig. \ref{fig2}(c) with readout phases $\theta=\pi~(-\pi)$, and in Fig. \ref{fig2}(d) with readout phases $\theta=3\pi/4~(-\pi/4)$, giving a readout order of 2-1-3. Additionally, we measured the readout fidelity for the orders 1-2-3, 2-1-3, and 3-2-1, as summarized in Table \ref{tab1}. All measured fidelities exceeded 99\%, demonstrating the high reliability of the random access storage process. {\color{black} To our knowledge, this is the first demonstration of RAQM for time-bin encoded qubits, which are particularly are particularly robust in long-distance fiber transmission  \cite{marcikic2003long} }
In our current demonstration, the number of random-access storage channels is primarily constrained by the storage time and the relatively slow rise time of the AOD ($1.9$ \us). The integrated spin-wave quantum storage, as demonstrated in our recent work \cite{liu2025millisecond}, can significantly extend both the storage time and the number of accessible channels.
%Storage efficiency, ranging from $2.9\%$ to $18.7\%$, depends on the readout order (see Supplemental Materials).

\begin{table}[h]
    \centering
    \caption{Storage fidelity of time-bin qubits $Q_1:(\ket{e}-i\ket{l})/\sqrt{2},~ Q_2: (\ket{e}+i\ket{l})/\sqrt{2}$ and $Q_3: (\ket{e}+\e^{i 3\pi/4}\ket{l})/\sqrt{2}$ with readout arranged in various orders.}
    \label{tab1}
    \begin{tabular*}{0.8\linewidth}{cccc}
        \hline\hline
        \textbf{Order} & $F_{Q_1}$&$F_{Q_2}$ & $F_{Q_3}$ \\
        \hline
        1-2-3 &$99.6\pm 0.1\%$ &$99.5\pm 0.1\%$ & $99.7\pm 0.1\%$ \\
        2-1-3 &$99.3\pm 0.1\%$ &$99.7\pm 0.1\%$ & $99.6\pm 0.1\%$ \\
        3-2-1 &$99.4\pm 0.2\%$ &$99.7\pm 0.1\%$ & $99.1\pm 0.1\%$ \\
        \hline\hline
    \end{tabular*}
\end{table}

\subsection*{Quantum storage of high-dimensional spatial states}

% \textbf{Quantum storage of high-dimensional spatial states.}
High-dimensional quantum information (qudits) \cite{erhard2020advances} has emerged as an indispensable resource in quantum networks, offering several advantages. These include higher information capacity \cite{bechmann2000quantum} with increased resistance to noise \cite{cerf2002security} in quantum communication, and the ability to simplify quantum logic \cite{lanyon2009simplifying} while enhancing fault tolerance \cite{campbell2014enhanced} in quantum computing. However, previous implementations of integrated photonic quantum memories have been limited to the storage of two-dimensional information \cite{saglamyurek2011broadband,zhong2017nanophotonic,zhou2023photonic,seri2019quantum,meng2026efficient,liu2025millisecond,liu2022demand}. Our multichannel integrated quantum memory overcomes this limitation, providing an ideal platform for the storage of path-encoded high-dimensional quantum states.

\begin{figure}[!htbp]
\centering
\includegraphics[width=0.49\textwidth]{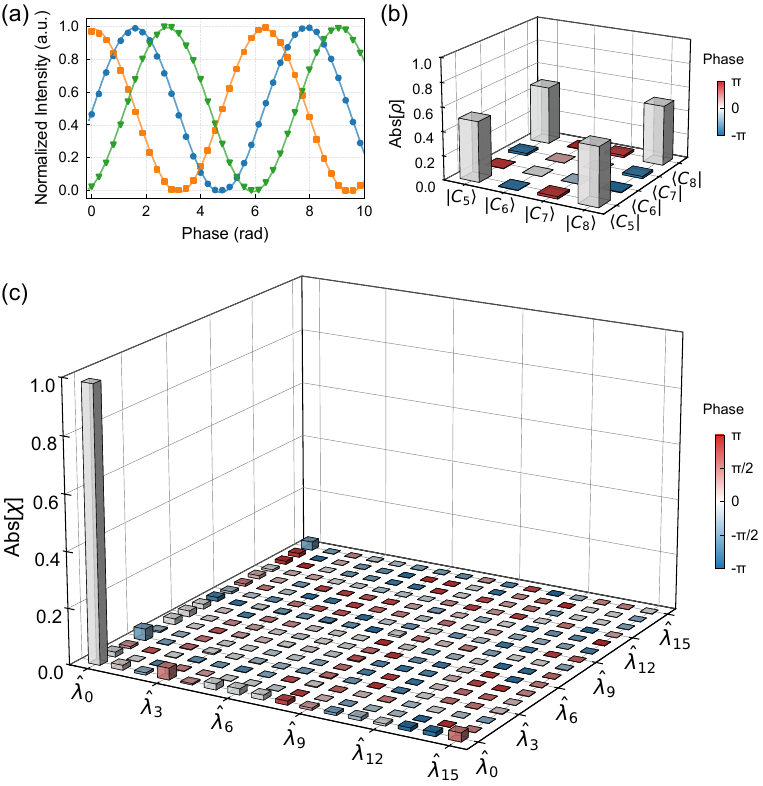}
\caption{Integrated quantum memory for four-dimensional path-encoded quantum information. (a) Interference fringes measured for three example input states: $(\ket{C_6}+\ket{C_7})/\sqrt{2}$~(blue circles), $(\ket{C_6}+i\ket{C_7})/\sqrt{2}$ (orange squares) and $(\ket{C_7}+\ket{C_8})/\sqrt{2}$ (green triangles). Data points are fitted with cosine functions (solid lines). (b) The reconstructed density matrix for input state  $(\ket{C_5}+\ket{C_8})/\sqrt{2}$ is shown, with a measured fidelity of $98.5\pm 0.6\%$. (c) The reconstructed process matrix $\chi$ for four-dimensional quantum storage is presented. The basis operators include $\hat{\lambda}_0=I_{4\times 4}$, the identify operator, and $\hat{\lambda}_{i},\ (1\leq i\leq 15)$, which are the generators of the SU(4) group. For (b) and (c), the height of the bar indicates the absolute value of the matrix elements, while the color represents the phase. }\label{fig3} 
\end{figure}

\begin{figure}[!htbp]
\centering
\includegraphics[width=0.49\textwidth]{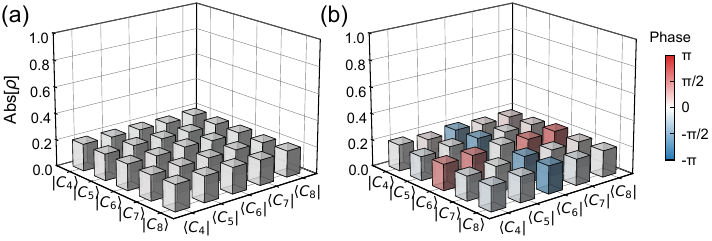}
\caption{Integrated quantum memory for five-dimensional quantum states. The reconstructed density matrix for outputs with inputs of $\ket{\psi_{1}}=(\ket{C_5}+\ket{C_6}+\ket{C_7}+\ket{C_8}+\ket{C_9})/\sqrt{5}$ (a) and $\ket{\psi_{2}}=(\ket{C_5}+\ket{C_6}+i\ket{C_7}+\ket{C_8}+\ket{C_9})/\sqrt{5}$ (b), with measured fidelities of $97.9\pm 0.5 \%$ and $96.1\pm 0.5 \%$, respectively. The height of the bar represents the absolute value of matrix elements while the color denotes the phase.}\label{fig4}
\end{figure}

In our experiments, photons can be coherently encoded into or projected onto $d$-dimensional path-encoded quantum states $\ket{\psi}=\sum_{n=1}^{d} a_n\ket{C_n}$, where $C_n$ denotes the memory channel $n$ and $a_n\in\mathbb{C}$ represents the complex superposition coefficient. These coefficients are precisely controlled through adjusting the amplitudes and phases of the RF signals driving AODs \cite{lan2009multiplexed,pu2017experimental}. As shown in Fig. \ref{fig3}(a), interference fringes for distinct input states demonstrate preserved inter-path coherence, with phase control implemented through RF phase adjustments to the read AOD. To perform four-dimensional quantum storage, we select channels 5, 6, 7, and 8 to prepare an AFC structure with a comb spacing of $2$ MHz. The input states are $\ket{\psi_{in}}=\ket{C_n},~(\ket{C_n}+\ket{C_m})/\sqrt{2}$, and $(\ket{C_n}+i\ket{C_m})/\sqrt{2}$, with mean photon number $\mu=0.98$ per pulse, where $5\leq n<m\leq 8$. For the $2^{nd}$ SMAFC echo retrieved at 1 \us, we perform four-dimensional quantum state tomography (QST) using the bases: $\ket{C_n},~(\ket{C_n}+\ket{C_m})/\sqrt{2}$, and $(\ket{C_n}+i\ket{C_m})/\sqrt{2}$, where $5\leq n<m\leq 8$. As an example, Fig. \ref{fig3}(b) displays the reconstructed density matrix of the retrieved state when the input is prepared as $(\ket{C_5}+\ket{C_8})/\sqrt{2}$. To fully characterize the multichannel memory performance for four-dimensional storage, we implement four-dimensional quantum process tomography (QPT) \cite{o2004quantum} (see Sec. 7 in Supplemental Materials). The reconstructed process matrix $\chi$ shown in Fig. \ref{fig3}(c) demonstrates a process fidelity of $F_\chi=97.9\pm 0.2 \%$ with respect to the identity operator. {\color{black} Previous demonstrations of high-dimensional quantum memory \cite{dong2023highly,li2020high,tiranov2017quantification} have certified high dimensionality mainly by showing a Schmidt number larger than two. However,
this criterion—especially in the presence of non-ideal state fidelity—does not demonstrate that a memory possesses a higher information capacity than a single qubit. Through quantum process tomography and optimization algorithms (see Sec. 8 in Supplemental Materials), we calculate a quantum channel capacity of $C=1.932\pm0.007$ for our integrated quantum memory—significantly surpassing the single-qubit limit of $C = 1)$. This result provides an information-theoretic certification that our storage device, by leveraging high‑dimensional encoding, achieves higher information density per photon. It thereby establishes a concrete performance benchmark for building high-dimensional quantum networks.}

We further demonstrate five-dimensional quantum state storage using path-superposition states $\ket{\psi_{1}}=(\ket{C_5}+\ket{C_6}+\ket{C_7}+\ket{C_8}+\ket{C_9})/\sqrt{5}$ and $\ket{\psi_{2}}=(\ket{C_5}+\ket{C_6}+i\ket{C_7}+\ket{C_8}+\ket{C_9})/\sqrt{5}$ encoded across channels 5, 6, 7, 8, and 9, with an input intensity of $\mu=0.38$. The measured storage efficiencies for states $\ket{\psi_1}$ and $\ket{\psi_2}$ are $30.2\pm 0.3\%$ and $31.4\pm 0.4\%$, respectively, for the $2^{nd}$ SMAFC echo retrieved at 1 \us. We perform quantum state tomography on the readout using the complete measurement basis set $\{\ket{C_n},~ (\ket{C_n}+\ket{C_m})/\sqrt{2},~ (\ket{C_n}+i\ket{C_m})/\sqrt{2},~ (5\leq n< m\leq 9)\}$. The reconstructed density matrices $\rho_1$ and $\rho_2$ corresponding input states $\ket{\psi_1}$ and $\ket{\psi_2}$ are shown in Fig. \ref{fig4}(a,b), demonstrating state fidelities of $ F_1=97.9\pm 0.5\%$ and $F_2=96.1\pm 0.5 \%$, respectively. These values significantly exceed the classical fidelity bound of $48\%$ (see Sec. 9 in Supplementary Materials for details) by more than 96 standard deviations, thereby demonstrating the reliable storage of qudits with $d=5$. The storage dimensionality of our device is intrinsically scalable: it scales directly with the number of individually addressable memory channels.  {\color{black}The number of channels can be scaled up to hundreds via two-dimensional waveguide arrays and crossed AOD pairs for addressing.}

\section*{Discussion}

In summary, we have developed an 11-channel integrated quantum memory based on laser-written waveguide arrays fabricated in a \euyso~crystal. Individual channels are addressed via AODs and independently controlled using on‑chip electrode arrays, with negligible inter‑channel crosstalk. {\color{black}This multichannel architecture introduces new functionalities to integrated photonic memories, enabling RAQM for time-bin qubits with an average fidelity exceeding 99\%, as well as quantum storage of five-dimensional states with a fidelity above 96\%. To achieve full on‑chip  integration, the current free‑space coupling based on AODs can be replaced by directly bonding the device to a fiber array using an ultraviolet‑curable adhesive \cite{rakonjac2022storage,liu2022demand,li2023handling}. {\color{black}The storage efficiency can be further improved to exceed 80\% by employing waveguide-based impedance-matched cavities, as recently demonstrated in our related work \cite{meng2026efficient}}}.

{\color{black}Looking ahead, independent optical control of each storage channel can also be realized via spin-wave storage protocols \cite{ma2021elimination,ortu2022storage}, which leverage long-lived spin coherence to support longer storage times.} Operation at zero magnetic field already allows millisecond‑scale storage \cite{liu2025millisecond}, suitable for multiplexed and high‑dimensional quantum repeaters. Under strong magnetic fields, storage times can be extended to minutes or even hours \cite{ma2021one,lv2025minute}, opening pathways toward high‑density, transportable quantum memories  \cite{gundoǧan2024time,bland2021quantum}.

% Looking forward, independent control of each storage channel can also be implemented optically via spin‑wave storage protocols \cite{ma2021elimination,ortu2022storage}, which support substantially longer storage times. Operation at zero magnetic field already allows millisecond‑scale storage \cite{liu2025millisecond}, suitable for multiplexed and high‑dimensional quantum repeaters. Under strong magnetic fields, storage times can be extended to minutes or even hours \cite{ma2021one,lv2025minute}, opening pathways toward high‑density, transportable quantum memories  \cite{gundoǧan2024time,bland2021quantum}.

\section*{Methods}

\subsection*{Multichannel addressing and encoding} 
The 11 memory channels are addressed using a write/read AODs and lenses with a focal length of 75 mm. As illustrated in Fig. \ref{fig1}(a), the AODs are positioned on the focal plane of the lenses, enabling precise horizontal beam deflection and coupling the beam into and out of each waveguide, which has a core diameter of 30 \um~core diameter. The RF sources for controlling the AODs are generated by a common arbitrary waveform generator (AWG). The frequency of the RF signal is scanned from 65.75 MHz to 104.25 MHz in steps of 3.85 MHz, allowing the AOD to deflect the beams into the 11 memory channels. The write AOD and the read AOD have opposite frequency offsets for each memory channel. As a result, the frequency differences between different channels are compensated after passing through the read AOD, facilitating interference measurements across multiple channels. Since all channels share the same optical setup, the relative phase between channels is inherently stable, enabling path encoding.

\subsection{Time-bin qubit analyzer}
The time-bin qubit analyzer functions as an unbalanced Mach-Zehnder interferometer, projecting time-bin qubits onto superposition states for analysis. It is implemented using an independent AFC memory with a storage time of $T=200$ ns, matching the time-bin separation interval. When a single photon passes through the AFC memory, it either transmits directly with probability $p_{0}$ or is emitted after a $200$ ns delay with probability $p_1$, labeled as early bin and late bin, respectively. By controlling the AFC structure to achieve $p_0=p_1$, the device effectively functions as a time-bin splitter. Additionally, the phase $\theta=2\pi\delta_f T$ between the early and late bin can be controlled by shifting the AFC frequency detuning  $\delta_f$ \cite{kutluer2019time}. 

After submission of this work, we note that multichannel qubit storage has been demonstrated in an array of bulk $\mathrm{Pr^{3+}:}\mathrm{Y}_2\mathrm{SiO}_5$ crystals with spin-wave storage \cite{teller2025quantum}, though without random-access capability and high-dimensional storage.

\section*{Declarations}
\textbf{Funding:} This work is supported by the Strategic Priority Research Program of the Chinese Academy of Sciences (No. XDB1740000), the Quantum Science and Technology-National Science and Technology Major Project (No. 2021ZD0301200, 2024ZD0301400), HFNL Self-Deployed Project (No. ZB2026020200), the National Natural Science Foundation of China (Nos. 12222411 and 12404572), the China Postdoctoral Science Foundation (2023M743400), and this work was partially carried out at the USTC Center for Micro and Nanoscale Research and Fabrication. Z.-Q. Z. acknowledges the support from the Youth Innovation Promotion Association CAS.
\\\\
\textbf{Authors’ contributions:} Z.-Q.Z. designed the experiment and supervised all aspects of this work; Z.-W.O. fabricated the device; Z.-W.O. and T.-X.Z. performed the experiment and analyzed the data; P.-J.L grew the crystal; X.-M.H. calculated the quantum channel capacity; Z.-W.O., T.-X.Z., X.-M.H and Z.-Q.Z. wrote the manuscript; Z.-Q.Z. and C.-F.L. supervised the project. All authors discussed the experimental procedures and results.
\\\\
\textbf{Data availability:} The main data supporting the finding of this study are available within the main text and supplementary information.
\\\\
\textbf{Competing interests:} The authors declare that they have no competing interests.

% After submission of this work, we note that multichannel qubit storage has been demonstrated in an array of bulk $\mathrm{Pr^{3+}:}\mathrm{Y}_2\mathrm{SiO}_5$ crystals with spin-wave storage \cite{teller2025quantum}, though without random-access capability and high-dimensional storage.
% Z.-Q.Z. designed the experiment and supervised all aspects of this work; Z.-W.O. fabricated the device; Z.-W.O. and T.-X.Z. performed the experiment and analyzed the data; P.-J.L grew the crystal; X.-M.H. calculated the quantum channel capacity; Z.-W.O., T.-X.Z., X.-M.H and Z.-Q.Z. wrote the manuscript; Z.-Q.Z. and C.-F.L. supervised the project. All authors discussed the experimental procedures and results.

%merlin.mbs apsrev4-1.bst 2010-07-25 4.21a (PWD, AO, DPC) hacked
%Control: key (0)
%Control: author (0) dotless jnrlst
%Control: editor formatted (1) identically to author
%Control: production of article title (0) allowed
%Control: page (1) range
%Control: year (0) verbatim
%Control: production of eprint (0) enabled
%

\clearpage
\newpage
\onecolumngrid

\renewcommand{\figurename}{Fig.}
\renewcommand{\tablename}{Table.}

\setcounter{table}{0}
\renewcommand{\thetable}{S\Roman{table}}
\setcounter{figure}{0}
\renewcommand{\thefigure}{S\arabic{figure}}
\setcounter{equation}{0}
\renewcommand{\theequation}{S\arabic{equation}}

\title{Supplementary Materials for\\ "Random access and high dimensional integrated quantum memory"}

\newpage \clearpage
\begin{center}
    {\Large\bfseries
    Supplementary Materials for\\ ``Random access and high dimensional integrated quantum memory''} \\[1.5em]

   {Zhong-Wen Ou\,$^{1,2,3,{\color{blue}*}}$, Tian-Xiang Zhu\,$^{1,2,3,{\color{blue}*}}$ Peng-Jun Liang\,$^{1,2,3,4}$ Xiao-Min Hu\,$^{1,2,3,4}$ Zong-Quan Zhou\,$^{1,2,3,4,{\color{blue}\dagger}}$ Chuan-Feng Li\,$^{1,2,3,4}$ and Guang-Can Guo\,$^{1,2,3,4}$} \\[0.5em]
    % \normalsize
    $^1$ \textit{\small Laboratory of Quantum Information, University of Science and Technology of China, Hefei 230026, China}\\
    $^{2}$ \textit{\small Anhui Province Key Laboratory of Quantum Network, University of Science and Technology of China, Hefei 230026, China}\\
    $^3$ \textit{\small CAS Center For Excellence in Quantum Information and Quantum Physics, University of Science and Technology of China, Hefei 230026, China}\\
    $^{4}$ \textit{\small Hefei National Laboratory, University of Science and Technology of China, Hefei 230026, China}
\end{center}

\newpage

\subsection{1. Details of the memory device}\label{sm_sec1}
\begin{figure}[!ht]
    \centering
    \includegraphics[width=\linewidth]{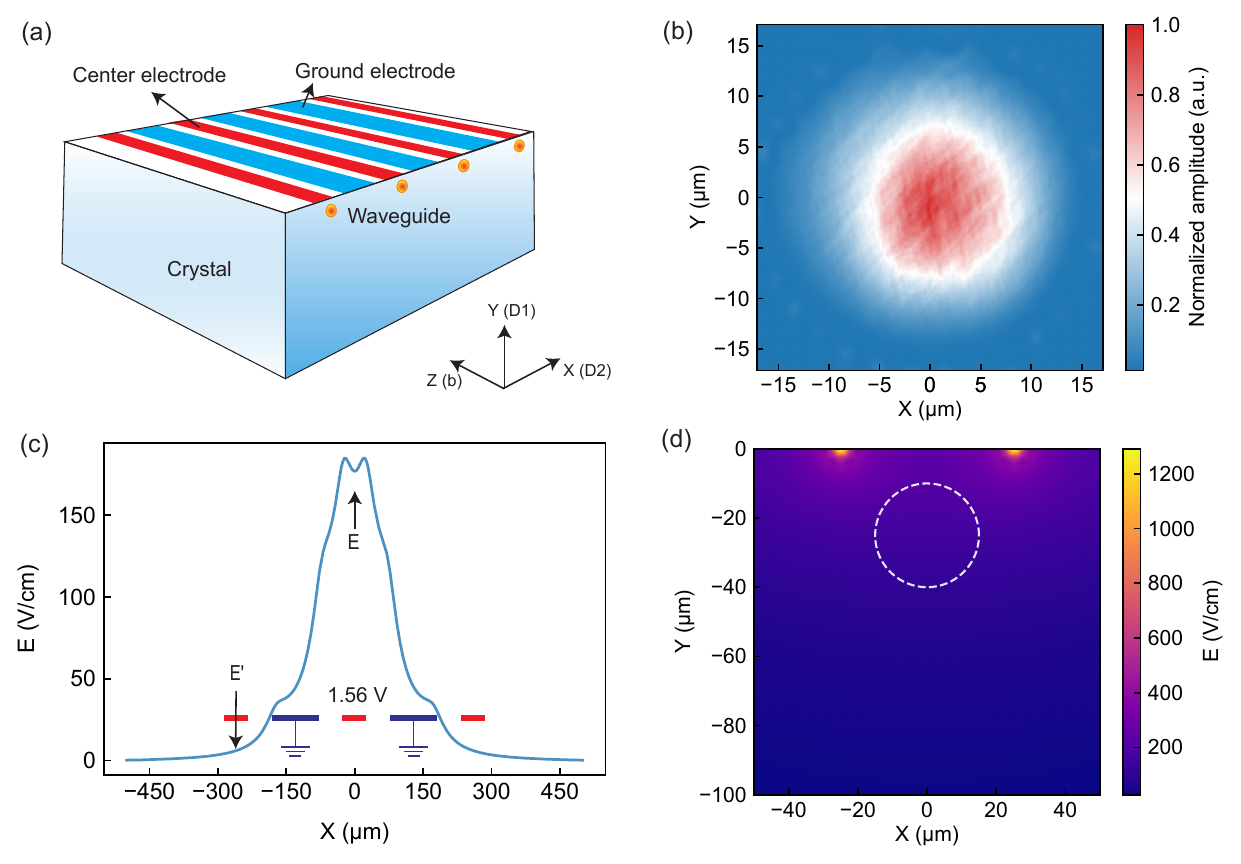}
    \caption{Multichannel memory device. (a) Schematic diagram of the memory structure. A waveguide array is aligned along the crystal's b axis with an inter-waveguide spacing of $250$ \um. The central electrode is positioned directly above the waveguide, with its center offset by $25$ \um~from the waveguide center. (b) Normalized mode field distribution at waveguide end-facet. (c) Electric field distribution along the X-axis at $25$ \um~depth below the electrode. The central electrode biased is $1.56$ V, where $E$ denotes the electric field magnitude beneath the central electrode, and $E'$ represents the residual field intensity beneath the adjacent electrode. (d) Electric field distribution in the XY cross-section; the dashed circle indicates the optical waveguide location.}
    \label{sm_fig1}
\end{figure}

The device architecture is illustrated in Fig. \ref{sm_fig1}(a). Waveguide arrays are sculpted with a femtosecond-laser micromachining (FLM) platform (WOPhotonics, Lithuania). The laser parameters include a wavelength of $1030$ nm, a repetition rate of $201.9$ kHz, a pulse duration of $219$ fs, and a pulse energy of $80$ nJ. The beam, polarized along the crystal's D2-axis, is focused 25 \um~below the crystal surface through a $100\times$, NA=0.7 objective and translated along the b-axis at a speed of $5$ mm/s. Each depressed cladding waveguide consists of $32$ parallel tracks arranged in a $30$-\um~radius circle. For signal mode with polarization along the D1 axis, we measure an optimized insertion loss of 0.46 dB (a device transmission of 90\%), which includes propagation, coupling, and Fresnel losses. {\color{black} These waveguides operate in single-mode at 580 nm, exhibiting a near-Gaussian guided mode with a fitted full width at half maximum (FWHM) of $14.6\times 15.1$ \um~(Fig. \ref{sm_fig1}(b)). Their single‑mode cutoff bandwidth is approximately 100 nm, supporting broadband pulse transmission.}
% The guided mode is nearly Gaussian, with fitted full width at half maximum (FWHM) of $14.6\times 15.1$ \um~(Fig. \ref{sm_fig1}(b)). Eleven waveguides are fabricated in the crystal with a spacing of $250$ \um. 
The absorption of $^{151}$Eu$^{3+}$ ion inside the optical waveguide is measured as $\alpha=4.3/$cm with a broadening of $1.46$ GHz, closely matching bulk values ($\alpha=4.6/$cm, $1.41$ GHz). After waveguide inscription, electrode arrays are patterned on the crystal surface via lift-off lithography. 

%insertion loss, including transmission loss and coupling loss, each below 0.5 dB/cm.
% Each depressed cladding waveguide consists of $32$ parallel tracks arranged in a $30$-\um~radius circle, yielding single-mode propagation (D1-polarization) with $>70\%$ (Fig. \ref{sm_fig3}(a)).

Extending the SMAFC protocol from a single channel to a multichannel integrated architecture presents two primary challenges: suppressing electrical crosstalk between adjacent channels and unavoidable inhomogeneity of the electric fields. To address these, we designed a tailored electrode geometry: each central electrode (width =$50$ \um) is aligned directly above a waveguide and flanked by $100$-\um~wide ground electrodes placed $125$ \um~away. This layout effectively minimizes inter‑electrode crosstalk. Applying a bias of 1.56 V to the central electrode yields the simulated field profiles shown in Figs. \ref{sm_fig1}(c,d). The resulting field at the waveguide center is $E=176.8$ V/cm, with a residual crosstalk field $E'=7.0$ V/cm$\approx 0.04 E$. As demonstrated in Fig. \ref{sm_fig5}, this residual crosstalk introduces less than 1\% variation in the storage efficiency of neighboring channels, confirming that electrical crosstalk is negligible in our architecture. Furthermore, here we demonstrate that, even though the electric field distribution within each waveguide is inhomogeneous, it does not compromise the performance of the SMAFC protocol. In Section 3, we provide a detailed model that incorporates a realistic Gaussian electric field distribution, showing that the suppression of the first‑order echo and the retrieval of higher‑order echoes remain fully effective under these practical conditions. %{\color{blue} In addition, the waveguide spacing is set to 250 $\upmu$m, primarily to suppress residual electric fields, which constrains the number of addressable memory channels to 11. In the future, spin‑wave storage protocols could eliminate the need for electrodes, enabling more channels and paving the way towards large‑scale integrated storage in two‑dimensional waveguide arrays.}

\subsection*{2. Details of experiment setup}\label{sm_sec2}
\begin{figure}[h]
    \centering
    \includegraphics[width=1\linewidth]{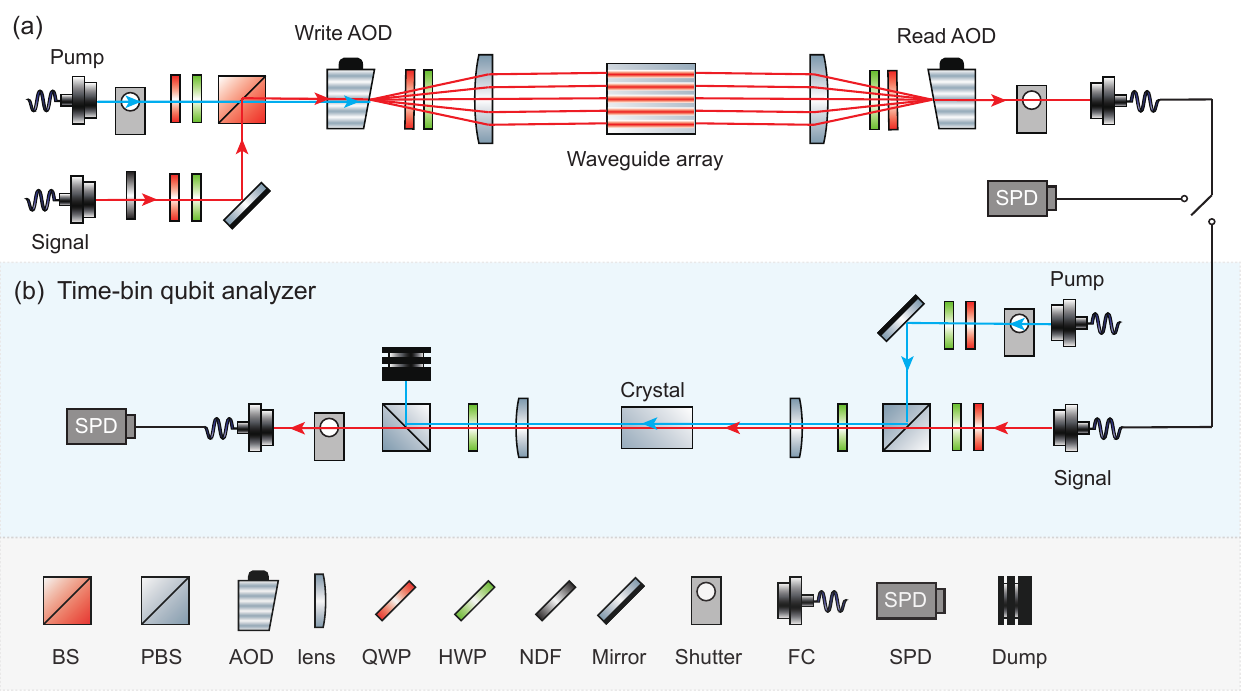}
    \caption{Detailed experimental optical setup. The optical configuration comprises two subsystems: multichannel storage unit (a) and time-bin qubit analyzer unit (b). (a), The signal beam is attenuated to single-photon level via a neutral density filters (NDF) and combined with pump beam through a $90:10$ (R:T) beam splitter (BS). The polarization control is achieved using a half-wave plate (HWP) and a quarter-wave plate (QWP). The combined beam is addressed to 11 memory channels (only 5 channels are schematically shown) through a write acoustic optical deflector (AOD) and a lens with 75 mm focal length.  Subsequent readout paths follow similar configurations, directing the output to either a single-photon detector (SPD) or a time-bin qubit analyzer. (b), The signal and pump beams are combined via a polarizing beam splitter (PBS), with polarization controlled by a HWP and a QWP. A 4f imaging system focuses the beam into the \euyso~crystal. The readout signal and the transmitted pump beams are separated by the PBS. Optical shutters are implemented throughout the system to protect SPD. Fiber collimator (FC).}
    \label{sm_fig2}
\end{figure}

As shown in Fig. \ref{sm_fig2}(a), the memory device is cooled to approximately $3.2$ K in a closed-cycle cryostat (Montana Instruments). The $580$ nm laser (Precilasers) is generated by sum-frequency of a $1537$ nm laser and a $932.5$ nm laser. The laser linewidth after external locking is approximately 0.17 kHz. The signal and pump beams are modulated using double-pass acousto-optic modulators (AOMs) driven by an arbitrary waveform generator (AWG, Zurich Instruments). This AWG also controls the Write/Read AODs (Brimrose) and electrode arrays. 
%The optical coherence time is measured as $383$ \us~via two-pulse photon echo.

%Polarization alignment along the crystal's D1 axis is optimized via a HWP and a QWP. Signal polarization is aligned along crystal's D1 axis using a HWP.

The 11 memory channels are addressed using a write/read AODs and lenses with a focal length of 75 mm. As illustrated in Fig. \ref{sm_fig2}(a), the AODs are positioned on the focal plane of the lenses, enabling precise horizontal beam deflection and coupling the beam into and out of each waveguide, which has a core diameter of 30 \um~core diameter. The RF sources for controlling the AODs are generated by a common AWG. The frequency of the RF signal is scanned from 65.75 MHz to 104.25 MHz in steps of 3.85 MHz, allowing the AOD to deflect the beams into the 11 memory channels. The write AOD and the read AOD have opposite frequency offsets for each memory channel. As a result, the frequency differences between different channels are compensated after passing through the read AOD, facilitating interference measurements across multiple channels. Since all channels share the same optical setup, the relative phase between channels is inherently stable, enabling path encoding.

The time-bin qubit analyzer shown in Fig. \ref{sm_fig2}(b) functions as an unbalanced Mach-Zehnder interferometer, projecting time-bin qubits onto superposition states for analysis. It is implemented using an independent AFC memory with a storage time of $T=200$ ns, matching the time-bin separation interval. The AFC structure is prepared using a pump beam. When a single photon passes through the AFC memory, it either transmits directly with probability $p_{0}$ or is emitted after a $200$ ns delay with probability $p_1$, labeled as early bin and late bin, respectively. By controlling the AFC structure to achieve $p_0=p_1$, the device effectively functions as a time-bin splitter. Additionally, the phase $\theta=2\pi\delta_f T$ between the early and late bin can be controlled by shifting the AFC frequency detuning  $\delta_f$  \cite{kutluer2019time}. 

% The time-bin qubit analyzer shown in Fig. \ref{sm_fig2}(b) characterizes the fidelity of stored time-bin qubits. The pump beam initializes and prepares the AFC structure with a comb spacing of $500$ kHz in the \euyso~crystal. In this crystal, the signal input splits into two components: a transmitted portion and a 1$^{st}$-order AFC echo emits at $T=200$~ns. For time-bin qubit analyzer operation, the comb structure must be adjusted to maintain amplitude balance between the transmitted component and the AFC echo. Spectral shifting of the AFC enables control over their relative phase $\theta=2\pi \delta_f T$, thereby modifying the readout phase, where $\delta_f$ denotes the frequency detuning.

\begin{figure}[h]
    \centering
    \includegraphics[width=1\linewidth]{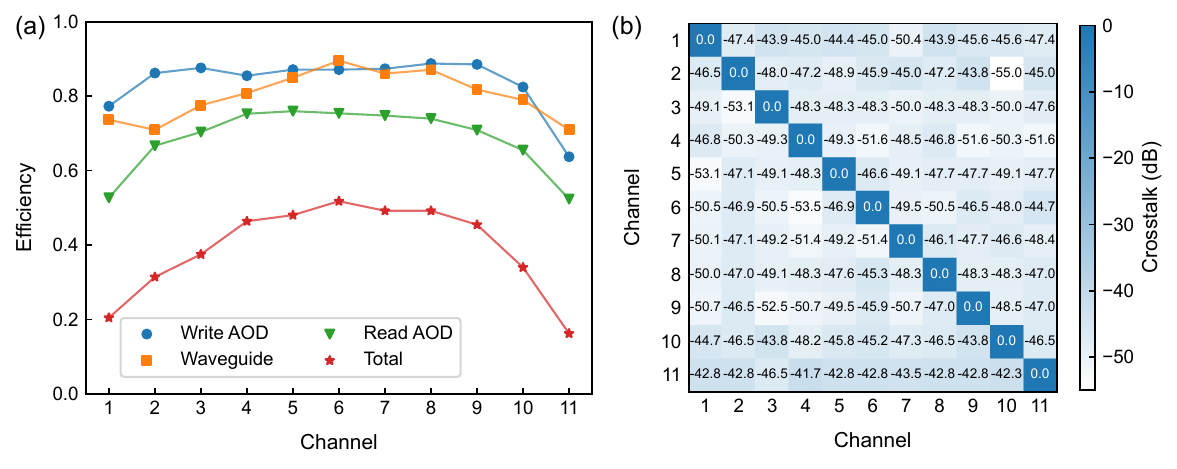}
    \caption{(a) Components efficiency for individual storage channels. Total efficiency is defined as the ratio of the output after final single-mode collection to the input prior to injection into the write AOD.  (b) Inter-channel crosstalk performance. }
    \label{sm_fig3}
\end{figure}

We present a detailed characterization of component efficiency within the storage unit, as illustrated in Fig. \ref{sm_fig3}(a). The total optical path efficiency across different memory channels ranges from $16\%$ to $52\%$. Optical path efficiency is relatively low for channels located on both sides. This reduction primarily results from two factors: losses due to AOD diffraction, and mode mismatch during waveguide coupling caused by spherical lens aberration. The total transmission of the eleven waveguides ranges from $71\%$ to $90\%$. Notably, the transmission of some waveguides exceeds the $84\%$ transmittance of the uncoated crystal itself. This enhancement is attributed to the waveguide cavity effect since the crystal end face provides a reflectivity of approximately $8\%$.

The optical crosstalk between storage channels is defined as $C_{ij}=10\ln (n_{ij}/n_{ii})$, where $n_{ij}$ represents the photon counts detected at the output of the $j$-th channel when photons are input into the $i$-th channel. Our measurements indicate that the optical crosstalk between storage channels remains below $-40$ dB. This excellent isolation results from the wide waveguide spacing as well as the effective photon confinement within the waveguide structure. Additionally, in Fig. \ref{sm_fig5}(b) in Section 3, we present a quantitative measurement of electrical crosstalk of our device.

\subsection*{3. SMAFC storage protocol}\label{sm_sec3}

The SMAFC protocol enables on-demand quantum storage by introducing an electric field to manipulate atomic coherence  \cite{horvath2021noise,liu2020demand,zhu2022demand}. Due to the space-group symmetry of \yso~crystal, applying an electric field along the crystal's D1 axis results in two distinct classes of ions experiencing the same Stark frequency shifts of magnitude, but in opposite directions. {\color{black} For a spatially distributed electric field $\mathbf{E}(x,y)=\mathbf{E}_0+\delta \mathbf{E}(x,y)$, where $\mathbf{E}_0$ is constant and $\delta\mathbf{E}$ is a small variation, the Stark shift is given by $\Omega(x,y)=(\boldsymbol{\mu}_g-\boldsymbol{\mu}_e)\cdot \mathbf{E}/h=\Omega_0+\delta \Omega$, where $\boldsymbol{\mu}_g$ and $\boldsymbol{\mu}_e$ denote the electric dipole moments of the ground and excited states, respectively, and $h$ is Planck's constant. After the photon absorption at time t = 0, the state of an AFC can be described as a collective excitation of the form \cite{horvath2021noise}
\begin{align}
    |\Psi(t)\rangle=\frac{1}{2\sqrt{M}} \sum_{l=0}^{M-1}\mathrm{e}^{i\omega_l t}\left[\int \mathrm{e}^{i 2\pi \Omega(x,y)\tau}|\psi_l^+(x,y)\rangle  \, \mathrm{d}x  \mathrm{d} y+\int \mathrm{e}^{-i 2\pi \Omega(x,y)\tau }|\psi_l^-(x,y)\rangle  \, \mathrm{d} x \mathrm{d} y \right], \label{eq10}
\end{align}
here, $M$ is the number of peaks in the AFC, $\omega_l=2\pi \Delta l$ with $\Delta$ is the separation between the AFC peaks, and $\tau$ is the duration of the electric-field pulse. $|\psi_l^+(x,y)\rangle$ and $|\psi_l^-(x,y)\rangle$ are the wave functions of the positive and negative electrically inequivalent ion classes, with the former defined as
\begin{align}
|\psi_l^\pm(x,y)\rangle=\frac{1}{\sqrt{N_l^\pm}}\sum_{j=1}^{N_l^\pm}c_{lj}^\pm(x,y)\mathrm{e}^{2\pi i \delta_{lj}(x,y)t} \mathrm{e}^{-i k z_{lj}^\pm(x,y)}|g_1\cdots e_j\cdots g_{N_l^\pm}\rangle, \label{eq11}
\end{align}
where $N_l^\pm$ is the number of ions in the $l$-th peak that experience a positive/negative frequency shift due to the electric field,  $c_{lj}^\pm $ is the probability amplitude for the input photon to excite the $j$-th ion in that peak,  $\delta_{lj}$ is the frequency detuning of the $j$-th ion from the peak’s central frequency $\omega_l$, $k$ is the photon wave vector, and $z_{lj}^\pm$ is the position of the $j$-th ion along the propagation direction. At time $t=n/\Delta$, the echo intensity as a function of the electric-pulse duration $\tau$ can be derived as
\begin{equation}
\begin{aligned}
     I_{echo}=  &\propto \left|\left|\int \,  \left[\mathrm{e}^{i 2 \pi \tau \Omega(x,y)}+\mathrm{e}^{-i 2 \pi \tau \Omega(x,y)}\right] \mathrm{d} x  \, \mathrm{d} y\right|\right|^2 \\
    &= \left|\left|\int \,\cos \left[2\pi \tau(\Omega_0+\delta \Omega)\right] \mathrm{d} x  \, \mathrm{d} y\right|\right|^2 \\
    &\propto \left|\left|\mathrm{Re} (\mathrm{e}^{i 2\pi \tau \Omega_0 }\langle\mathrm{e}^{i 2\pi \tau\delta \Omega }\rangle)\right|\right|^2. \label{eq12}
\end{aligned}
\end{equation}

For a uniform electric field $(\delta \Omega=0)$, yield $I_{echo}\propto \cos^2(2\pi \Omega_0 \tau)$. If $\delta\Omega$ follows a Gaussian distribution with zero mean and variance $\sigma_\Omega$, i.e., $\delta \Omega\sim N(0,\sigma_\Omega)$, then $\langle\mathrm{e}^{i 2\pi \tau \delta\Omega }\rangle=\mathrm{e}^{-2\pi^2 \sigma_\Omega^2 \tau^2}$, and the expression of Eq. \ref{eq12} becomes
\begin{align}
    I_{echo}\propto\cos^2 (2\pi \Omega_0 \tau)\mathrm{e}^{-4\pi^2 \sigma_\Omega^2 \tau^2}. \label{eq13}
\end{align}
Therefore, applying an electrical pulse with duration $\tau=1/(4\Omega_0)$ during the time interval $[0,1/\Delta]$ effectively suppresses the emission of the standard AFC echo, an effect that is independent of the spatial uniformity of the electric field.} Subsequently, a second pulse of negative polarity applied during the time interval  $[(n-1)/\Delta,n/\Delta]$ induces photon re-emission at $t=n /\Delta$, generating the \nth -order SMAFC echo. The storage efficiency of SMAFC is given by
\begin{equation}
    \eta=\tilde{d}^2 {\rm e} ^{-\tilde{d}}{\rm e}^{-t^2\tilde{\gamma}^2}, \label{eq1}
\end{equation}
where $\tilde{\gamma}$ is related to the comb FWHM by $\tilde{\gamma}=2\pi \gamma/\sqrt{8\ln 2}$, $\tilde{d}$ is the effective absorption depth of the comb defined as $\tilde{d}=\frac{d}{F}\sqrt{\frac{\pi}{4\ln 2}}$, $F$ is the the comb finesse, and $d$ is the peak absorption of the combs \cite{horvath2021noise}.

% The SMAFC protocol enables on-demand quantum storage by introducing an electric field to manipulate atomic coherence  \cite{horvath2021noise,liu2020demand,zhu2022demand}. Due to the space-group symmetry of \yso~crystal, applying an electric field along the crystal's D1 axis results in two distinct classes of ions experiencing the same Stark frequency shifts of magnitude, but in opposite directions.  The shift magnitude is given by $\Omega=(\boldsymbol{\mu}_g-\boldsymbol{\mu}_e)\cdot \mathbf{E}/h$, where $\boldsymbol{\mu}_g$ and $\boldsymbol{\mu}_e$ denote the electric dipole moments of the ground and excited states, respectively, $\mathbf{E}$ is the applied electric field, and $h$ is Planck's constant. After the photon absorption at $t=0$, applying an electrical pulse with duration $T=1/(4\Omega)$ during the time interval $[0,1/\Delta]$ effectively suppresses the emission of the standard AFC echo. Subsequently, a second pulse of negative polarity applied during the time interval  $[(n-1)/\Delta,n/\Delta]$ induces photon re-emission at $t=n /\Delta$, generating the \nth -order SMAFC echo. The storage efficiency of SMAFC is given by
% \begin{equation}
%     \eta=\tilde{d}^2 {\rm e} ^{-\tilde{d}}{\rm e}^{-t^2\tilde{\gamma}^2}, \label{eq1}
% \end{equation}
% where $\tilde{\gamma}$ is related to the comb FWHM by $\tilde{\gamma}=2\pi \gamma/\sqrt{8\ln 2}$, $\tilde{d}$ is the effective absorption depth of the comb defined as $\tilde{d}=\frac{d}{F}\sqrt{\frac{\pi}{4\ln 2}}$, $F$ is the the comb finesse, and $d$ is the peak absorption of the combs \cite{horvath2021noise}.

\begin{figure}[t]
    \centering
    \includegraphics[width=0.8\linewidth]{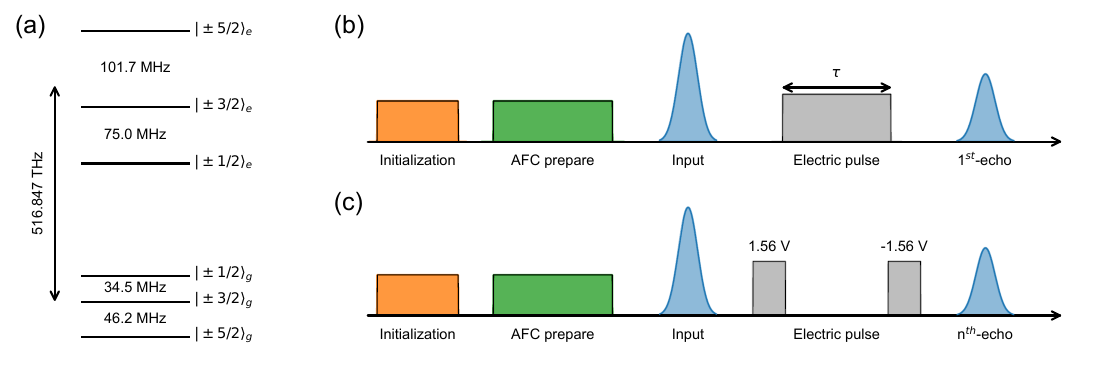}
    \caption{(a) Level structure for site-1 $^{151}$Eu$^{3+}$ ions at zero magnetic field. (a) Time sequence for Stark modulation measurements with variable modulation pulse duration $\tau$. (c) Time sequence for SMAFC storage protocol.}
    \label{sm_fig4}
\end{figure}

The energy level structure of the $^7F_0\to {^5D_0}$ transition of $^{151}$Eu$^{3+}$ ions is illustrated in Fig. \ref{sm_fig4}(a) and the time sequence for SMAFC storage is illustrated in Fig. \ref{sm_fig4}(c). During initialization, we optically pump ions within frequency ranges $[f_0+10.15~\mathrm{MHz},f_0+90.85 \mathrm{MHz}]$ and $[f_0-90.85 \mathrm{MHz},f_0-10.15 \mathrm{MHz}]$ to enhance the optical depth and bandwidth, creating an absorption profile centered at $f_0$ with  $20~\mathrm{MHz}$ bandwidth and with an absorption depth of $\sim 10.7$ \cite{zhu2022demand,liu2024nonlocal}. {\color{black} The current bandwidth is limited by the hyperfine level spacing of the lower-lying states in \(^{151}\mathrm{Eu}^{3+}\) ions; however, \(^{153}\mathrm{Eu}^{3+}\) ions could be utilized to achieve a larger bandwidth.} Subsequently, a parallel comb preparation scheme \cite{jobez2016towards} is used to generate an AFC with a comb tooth spacing of $\Delta=2~\mathrm{MHz}$ or $200~\mathrm{kHz}$. 

After the input photons are absorbed by the AFC structure, we apply an electric pulse with a duration of $50$ ns  and a voltage of $1.56$ V in the time interval $[0,1/\Delta]$, and a negative polarity pulse in the time interval $[(n-1)/\Delta,n/\Delta]$ to the electrodes, enabling the retrieval of the \nth -order echo at $t=n/\Delta$. 

\begin{figure}[h]
    \centering
    \includegraphics[width=\linewidth]{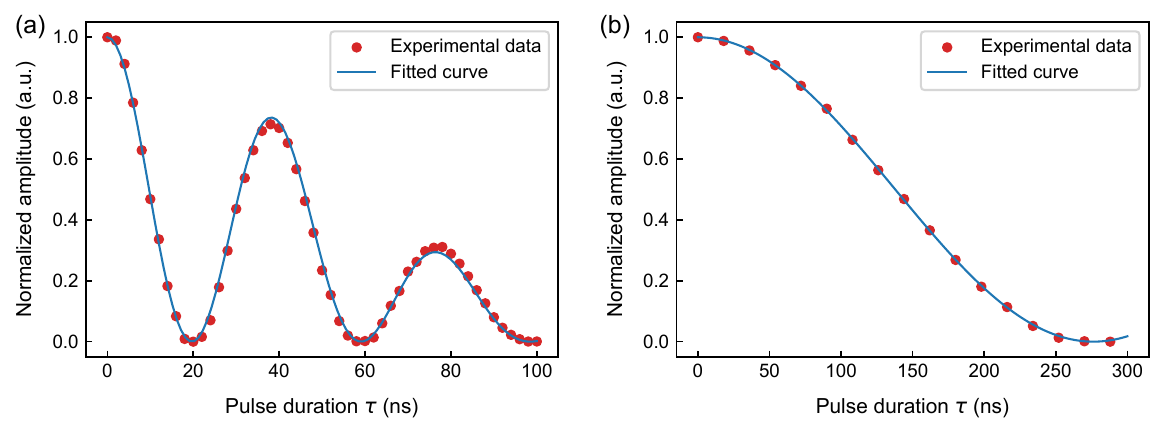}
    \caption{(a) Normalized amplitude of the 1$^{st}$-order AFC echo as a function of modulation pulse duration $\tau$, measured under $4$ V electrical pulses. (b) Normalized amplitude of the 1$^{st}$-order AFC
    echo versus modulation pulse duration $\tau$,  measured under $5$ V pulses applied to the electrode adjacent to the central electrode. For both (a) and (b), red markers denote experimental data;  blue solid lines represent theoretical fitted curves.}
    \label{sm_fig5}
\end{figure}

Before SMAFC storage implementation, we characterize the Stark modulation induced intereference of $^{151}$Eu$^{3+}$ ions using the sequence presented in Fig. \ref{sm_fig4}(b). The amplitude of the $1^{st}$-order echo is measured while varying electric pulse duration $\tau$ (Fig. \ref{sm_fig5}(a)). The echo amplitude exhibits damped oscillations with increasing $\tau$. For a uniform electric field, echo intensity follows: $I_{\text{echo}}\propto\cos^2(2\pi \Omega_0 \tau)$. Accounting for electric field inhomogeneity across the ensemble, the echo intensity satisfy Eq. \ref{eq13}, where $\sigma_\Omega$ characterizes the electric field inhomogeneity. Experimental data show excellent agreement with theoretical fits (Fig. \ref{sm_fig5}(a)), yielding: $\Omega_0=12.7$ MHz, $\sigma_\Omega=3.8$ MHz. The Stark coefficient is derived as: $\mu=28.0$ kHz/(V$\cdot$ cm$^{-1}$). This result is consistent with previously reported values \cite{macfarlane2014optical,liu2020demand}.

In order to further evaluate the impact of electrical crosstalk on the storage efficiency, an electric field is applied to the adjacent center electrode to perform the similar measurements on Stark modulation. The measured $1^{st}$-order AFC echo under this configuration is shown in Fig. \ref{sm_fig5}(b). Data fitting yields $\Omega'_0=0.9$ MHz, corresponding to a residual electric field $E'=0.05E$ ($E$ denotes the electric field of the central electrode), which is consistent with the electric field simulation results presented in Section 1. As a result, for the actual storage experiments (electric pulse duration of $50$ ns with a voltage of $1.56$ V), the efficiency attenuation due to crosstalk is estimated to be below $1\%$.

\begin{figure}[h]
    \centering
    \includegraphics[width=\linewidth]{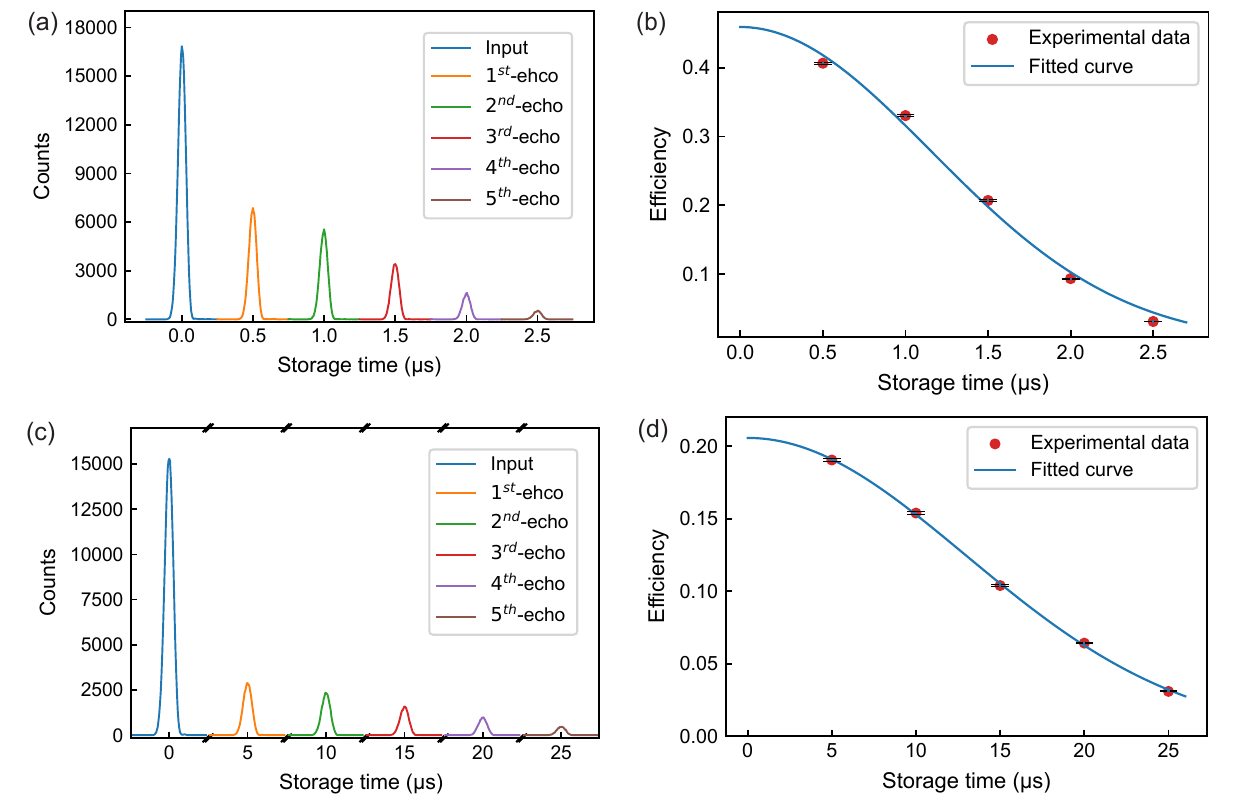}
    \caption{Photon counting histogram for SMAFC storage with $1/\Delta=0.5$ \us~(a) and $1/\Delta=5$ \us~(c). The SMAFC storage efficiency as a function of storage time for $1/\Delta=0.5$ \us~(b) and $1/\Delta=5$ \us~(d). For (b) and (d), red dots denote experimental data, and blue solid lines represent  theoretical fitted curves.}
    \label{sm_fig6}
\end{figure}

\begin{figure}[t]
    \centering
    \includegraphics[width=0.5\linewidth]{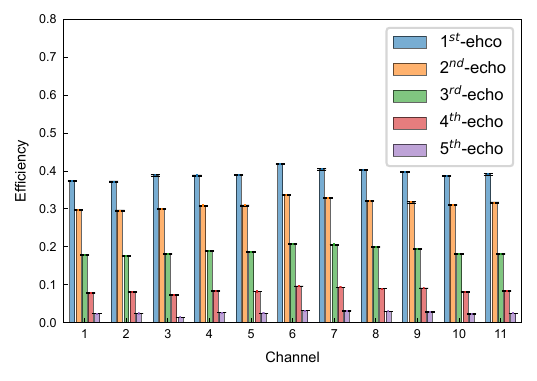}
    \caption{SMAFC storage efficiency across 11 memory channels with $\Delta=2$ MHz.}
    \label{sm_fig8}
\end{figure}

We prepare AFC with $\Delta=2$ MHz and $\Delta=200$ kHz, corresponding to SMAFC storage with step of $1/\Delta=0.5$ \us~ and $1/\Delta=5$~ \us, respectively. For $\Delta=2$ MHz, the preparation time of AFC is $2$ s. Fig. \ref{sm_fig6}(a) displays the photon counting histogram for SMAFC storage with a maximum storage time of $2.5$ \us. Fig. \ref{sm_fig6}(b) shows time-dependent storage efficiency. Fitting the data with Eq. \ref{eq1} yields the AFC finesse $F = 8.7$ and the absorption depth of the comb peaks $d = 10.6$. For $\Delta=200$ kHz, the preparation time of AFC is $5$ s. Fig. \ref{sm_fig6}(c) displays the photon counting histogram for SMAFC storage with a maximum storage time of $25$ \us. Fig. \ref{sm_fig6}(d) shows time-dependent storage efficiency with fitted $F = 9.8$ and $d = 5.7$. To comprehensively characterize the storage performance across all 11 channels, we implement SMAFC storage for each memory channel with $\Delta=2$ MHz. Storage efficiency versus time for all channels are shown in Fig. \ref{sm_fig8}. All channels exhibit nearly identical efficiency, demonstrating high uniformity across 11 memory channels.

% First, we address the target memory channel and prepare a 2 MHz AFC structure with a preparation time of 2 s. Executing $6\times 10^5$ write-read cycles using weak coherent light, with storage times dynamically adjusted through on-demand modification of electric pulse intervals, achieving maximum storage time of $2.5$ \us. Photon counting histograms are presented in main text Fig. 2(a). 

\newpage
\clearpage
\subsection*{4. Random access quantum memory}\label{sm_sec4}

% \begin{figure}[!b]
%     \centering
%     \includegraphics[width=\linewidth]{SM_Timebin random acess quantum memory.pdf}
%     \caption{Random access quantum storage of time-bin qubits with read-out in the order 1-2-3. (a) Photon counting histogram for time-bin qubit storage. $Q_1: \ket{e}/\ket{l}$ (blue solid / dashed), $Q_2: \ket{e}/\ket{l}$ (orange solid / dashed) and $Q_3: \ket{e}$ (green solid / dashed) are written and read-out in the order 1-2-3. (b) Interference histogram for qubits $Q_1,~Q_2,~Q_3$ prepared in $(\ket{e}+i\ket{l})/\sqrt{2}$. The read-out phases are set to $\theta=\pi/2$ (blue solid) and $\theta=-\pi/2$ (orange solid). Measured visibilities for the qubits are $97.3\pm 0.1\%,~ 98.9 \pm 0.1\%$, and $99.1 \pm 0.1\%$, respectively. (c) Interference histogram for qubits $Q_1,Q_2,Q_3$ prepared in $(\ket{e}+e^{i3\pi/4}\ket{l})/\sqrt{2}$. The read-out phases are set to $\theta=3\pi/4$ (green solid) and $\theta=-\pi/4$ (red solid). Measured visibilities for the qubits are $97.0\pm 0.1\%,~ 99.6 \pm 0.1\%$, and $98.8 \pm 0.1\%$, respectively.}
%     \label{sm_fig9}
% \end{figure}

\begin{figure}[h]
    \centering
    \includegraphics[width=\linewidth]{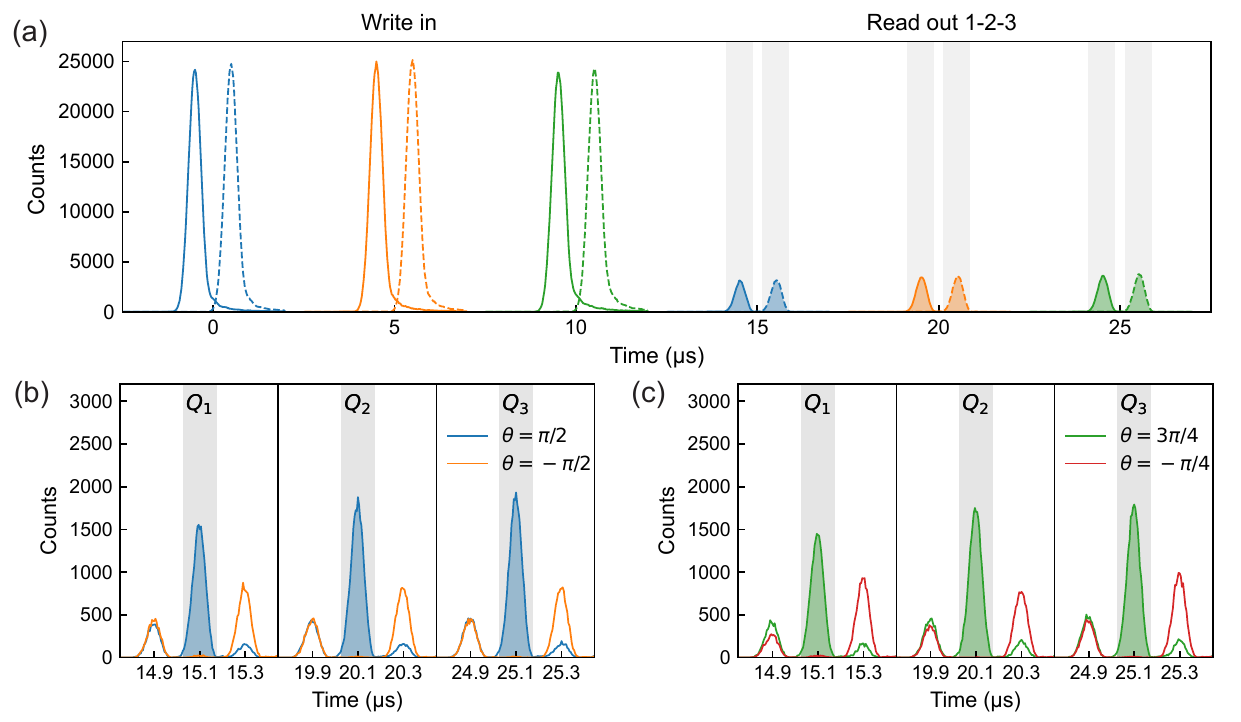}
    \caption{Random access quantum storage of time-bin qubits with read-out in the order 1-2-3. (a) Photon counting histogram for time-bin qubit storage. $Q_1: \ket{e}/\ket{l}$ (blue solid / dashed), $Q_2: \ket{e}/\ket{l}$ (orange solid / dashed) and $Q_3: \ket{e}$ (green solid / dashed) are written and read-out in the order 1-2-3. (b) Interference histogram for qubits $Q_1,~Q_2,~Q_3$ prepared in $(\ket{e}+i\ket{l})/\sqrt{2}$. The read-out phases are set to $\theta=\pi/2$ (blue solid) and $\theta=-\pi/2$ (orange solid). Measured visibilities for the qubits are $97.3\pm 0.1\%,~ 98.9 \pm 0.1\%$, and $99.1 \pm 0.1\%$, respectively. (c) Interference histogram for qubits $Q_1,Q_2,Q_3$ prepared in $(\ket{e}+e^{i3\pi/4}\ket{l})/\sqrt{2}$. The read-out phases are set to $\theta=3\pi/4$ (green solid) and $\theta=-\pi/4$ (red solid). Measured visibilities for the qubits are $97.0\pm 0.1\%,~ 99.6 \pm 0.1\%$, and $98.8 \pm 0.1\%$, respectively.}
    \label{sm_fig9}
\end{figure}

This section provide the details about random access quantum storage of multiple time-bin qubits. Time-bin qubits are prepared with a double-pass AOM. They are mapped into memory channels using the write AOD. Subsequently, the stored time-bin qubits can be read out in arbitrary order using a read AOD and a precisely controlled electrical pulse sequence. In addition to the data presented in the main text, here we provide another example with readout order of 1-2-3 in Fig. \ref{sm_fig9}. In Fig. \ref{sm_fig9}(a), qubits $Q_1:\ket{e}/\ket{l}$, $Q_2:\ket{e}/\ket{l}$, and $Q_3:\ket{e}/\ket{l}$ are sequentially written into the memory channels and retrieved in the order 1-2-3. The storage fidelity for the states $\ket{e}$ and $\ket{l}$ is defined as $F_{\ket{e}(\ket{l})} = \frac{S + N}{S + 2N}$, where $N$ represents the noise counts and $S$ the signal counts. When $Q_1,~Q_2$, and $Q_3$ are encoded in superposition states $(\ket{e}+i\ket{l})/\sqrt{2}$ and $(\ket{e}+e^{i3\pi/4}\ket{l})/\sqrt{2}$, the readout state is measured using the time-bin analyzer. The readout basis is defined as $(\ket{e}+e^{i\theta}\ket{l})/\sqrt{2}$, where $\theta$ denotes the read-out phase. The resulting interference are shown in Fig. \ref{sm_fig9}(b,c). The fidelity for superposition states is defined as $F = (V + 1)/2$, where $V$ denotes the measured interference visibility.
% There are small distortions for the measured input waveform which are resulted from slight bandwidth mismatch of the analyzing crystal.

\begin{table}[!h]
\caption{The fidelities of the three time-bin qubits after being retrieved in different orders.}\label{sm_tab1}
\begin{tabular*}{\textwidth}{@{\extracolsep\fill}lcccccc}
\hline\hline
\multicolumn{2}{@{}l@{}}{\textbf{Order}}
&$F_{\ket{e}}$ &$F_{\ket{l}}$ &$F_{(\ket{e}+i\ket{l})/\sqrt{2}}$ &$F_{(\ket{e}+e^{i3\pi/4}\ket{l})/\sqrt{2}}$ &$F_T$\\
\hline
&$Q_1$& $99.8\pm 0.1\%$ & $99.9\pm 0.1\%$ & $98.6\pm 0.1\%$ & $98.5\pm 0.1\%$ & $99.0\pm 0.1\%$ \\
1-2-3 &$Q_2$& $99.9\pm 0.1\%$ & $99.9\pm 0.1\%$ & $99.5\pm 0.1\%$ & $99.8\pm 0.1\%$ & $99.7\pm 0.1\%$ \\
 &$Q_3$ & $99.8\pm 0.1\%$ & $99.9\pm 0.1\%$ & $99.6\pm 0.1\%$ & $99.4\pm 0.1\%$ & $99.6\pm 0.1\%$\\
\hline
 &$Q_1$& $99.8\pm 0.1\%$ & $99.9\pm 0.1\%$ & $99.1\pm 0.1\%$ & $99.0\pm 0.1\%$ & $99.3\pm 0.1\%$\\
2-1-3 &$Q_2$& $99.8\pm 0.1\%$ & $99.9\pm 0.1\%$ & $98.6\pm 0.1\%$ & $99.0\pm 0.1\%$ & $99.1\pm 0.1\%$ \\
 &$Q_3$ & $99.8\pm 0.1\%$ & $99.9\pm 0.1\%$ & $99.6\pm 0.1\%$ & $99.5\pm 0.1\%$ & $99.7\pm 0.1\%$\\
\hline
 &$Q_1$& $99.8\pm 0.1\%$ & $99.9\pm 0.1\%$ & $99.3\pm 0.1\%$ & $99.0\pm 0.1\%$ & $99.4\pm 0.1\%$ \\
3-2-1 &$Q_2$& $99.8\pm 0.1\%$ & $99.9\pm 0.1\%$ & $99.8\pm 0.1\%$ & $99.6\pm 0.1\%$ & $99.8\pm 0.1\%$ \\
 &$Q_3$ & $99.7\pm 0.1\%$ & $99.9\pm 0.1\%$ & $99.3\pm 0.1\%$ & $98.6\pm 0.1\%$ & $99.2\pm 0.1\%$\\
\hline\hline
\end{tabular*}
\end{table}

Table \ref{sm_tab1} presents the complete fidelity data for storing the states $\ket{e},~\ket{l},~(\ket{e}+i\ket{l})/\sqrt{2}$, and $(\ket{e}+e^{i3\pi/4}\ket{l})/\sqrt{2}$, with read out orders 1-2-3, 2-1-3, and 3-2-1. The total fidelity is defined as $F_T = \frac{1}{3}F_{el} + \frac{2}{3}F_{\pm}$, where $F_{el} = {(F_{\ket{e}} + F_{\ket{l}})}/{2}$ and $F_{\pm} = ({F_{(\ket{e}+i\ket{l})/\sqrt{2}} + F_{(\ket{e}+e^{i3\pi/4}\ket{l})/\sqrt{2}}})/{2}$. The total fidelity exceeds $99\%$ for all readout sequences, which is far beyond the classical bound ($81.8\%$, see details in Section 9), demonstrating the excellent performance of the random access quantum memory. Table \ref{sm_tab2} lists storage efficiencies of time-bin qubits for different readout orders, ranging from $2.9\%$ to $18.7\%$.

\begin{table}[h]
    \centering
    \caption{Storage efficiencies for time-bin qubits $Q_1$, $Q_2$, and $Q_3$ with readout orders 1-2-3, 2-1-3, and 3-2-1. }
    \label{sm_tab2}
    \begin{tabular}{cccc}
        \hline\hline
        \textbf{Order} & $\eta_{Q_1}$&$\eta_{Q_2}$ & $\eta_{Q_3}$ \\
        \hline
        1-2-3 &$10.2\pm 0.1\%$ &$11.3\pm 0.1\%$ & $11.9\pm 0.1\%$ \\
        2-3-1 &$6.1\pm 0.1\%$&$15.2\pm 0.1\%$ & $11.6\pm 0.1\%$ \\
        3-2-1 &$18.7\pm 0.1\%$ &$10.5\pm 0.1\%$ & $2.9\pm 0.1\%$ \\
        \hline\hline
    \end{tabular}
\end{table}

\subsection*{5. Quantum storage for path-encoded qubits}\label{sm_sec5}
\begin{figure}[h]
    \centering
    \includegraphics[width=\linewidth]{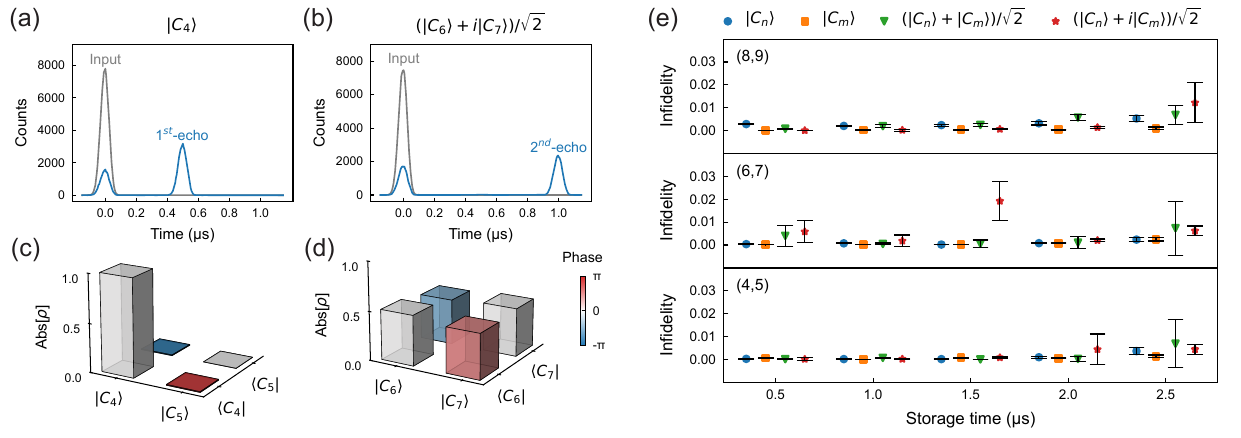}
    \caption{On-demand quantum storage of path-encoded qubits. (a,c) Path qubit $\ket{C_4}$ stored for $0.5$ \us. Reconstructed density matrix achieves a fidelity of $99.9\pm 0.1\%$. (b,d) Path qubit $(\ket{C_6}+i\ket{C_7})/\sqrt{2}$ stored for $1.0$ \us. Reconstructed density matrix achieves a fidelity of $99.9\pm 0.1\%$. (e) Measured storage infidelity for qubits encoded with memory channel pairs $(4,5),~(6,7)$ and $(8,9)$, where infidelity is defined as $1-F$. For (c) and (d), the height of the bar represents the absolute value of matrix elements, while the color denotes the phase.}
    \label{sm_fig10}
\end{figure}

Path-encoded qubits are prepared and measured via AODs. We dedicate the channel pairs $(n,m)$=(4,5), (6,7) and (8,9) to store the path qubits: $\ket{C_n},~(\ket{C_n}+\ket{C_m})/\sqrt{2}$ and $(\ket{C_n}+i\ket{C_m})/\sqrt{2}$, setting the mean photon number to $\mu=0.38$ per qubit. Fig. \ref{sm_fig10}(a,b) present photon-counting histograms for $\ket{C_4}$ and $(\ket{C_6}+\ket{C_7})/\sqrt{2}$, retrieved on demand after $0.5$~\us ~and $1.0$~\us, respectively. The corresponding reconstructed density matrices are shown in Fig. \ref{sm_fig10}(c,d). We further quantify the storage infidelity across all memory channel pairs for variable storage times (Fig. \ref{sm_fig10}(e)), demonstrating that all path qubits retain a fidelity exceeding $98\%$.

\subsection*{6. High-dimensional quantum state tomography}\label{sm_sec6}

Here we employ high-dimensional quantum state tomography (QST) for reconstructing the density matrix $\rho$ of qudit states \cite{james2001measurement}. For a $d$-dimensional quantum state, the density matrix $\rho_d$ admits the decomposition  
\begin{equation}
    \rho_d=\frac{1}{d}\sum_{i=0}^{d^2-1}r_i\hat{\lambda}_i,\label{eq2}
\end{equation}
where $\hat{\lambda}_0=I_{d\times d}$ and $\hat{\lambda}_i$ ($0<i\leq d^2-1$) denote the generators of the SU(d) group. The coefficients $r_i$ correspond to expectation values: $r_i=\langle \hat{\lambda}_i\rangle=\mathrm{Tr}(\rho\hat{\lambda}_i)$. Given a complete set of measurement bases $\{\hat{\mu}_i=|\psi_i\rangle\langle\psi_i|\}$, the projection operators can be expressed in terms of $\hat{\lambda_i}$ as: $\hat{\mu}_i=\sum_j A_i^j\hat{\lambda}_j$. For a d-dimensional quantum system, projective measurements yield outcomes: $n_i=N\mathrm{Tr}( \rho_d\hat{\mu})=N\sum_j A_i^j \mathrm{Tr}(\rho \hat{\lambda}_i)=N\sum_j A_{i}^j r_j$ ($ N$ is a normalization constant determinable from experimental data). The density matrix can formally be reconstructed through linear inversion: $\rho=N^{-1} \sum_{i,j}(A_i^j)^{-1} n_i \hat{\lambda}_j/d$. However, direct inversion may violate the positive semi definiteness of the density matrix (e.g., $\mathrm{Tr}(\rho)>0$). To ensure physicality, we employ maximum likelihood estimation (MLE) – introducing positivity and trace constraints to obtain the optimal density matrix $\hat{\rho}$.

The density matrix $\rho$ can be decomposed into the following form 
\begin{equation}
    \rho(t)=\hat{T}^{\dagger}(t)\hat{T}(t)/{\rm Tr}\{\hat{T}^\dagger(t)\hat{T}(t)\},\label{eq3}
\end{equation}
where $\hat{T}(t)$ denotes a parameterized matrix, it is convenient to adopt an upper triangular representation. The parameter set $t=\{t_1,t_2,...,t_{d^2}\}$ fully characterizes the matrix \cite{james2001measurement}. Crucially, this decomposition guarantees that $\rho$ satisfies semi-positive definiteness. The density matrix can be reconstructed by numerically minimizing the function:
\begin{equation}
    \mathcal{L}(t_1,t_2,...,t_{d^2})=\sum_{i=1}^{d^2} \frac{[N \langle \psi_i |\rho(t_1,t_2,...,t_{d^2})|\psi_i\rangle-n_i]^2}{2 N \langle \psi_i |\rho(t_1,t_2,...,t_{d^2})|\psi_i\rangle},\label{eq4}
\end{equation}

In our experimental configuration, the measurement bases are selected as: $\{\ket{C_n},~\{\ket{C_n}+\ket{C_m}/\sqrt{2},\quad (\ket{C_n}+i\ket{C_m})/\sqrt{2},\quad 1\leq n < m \leq d\}$. Photon counts for different projective measurements are recorded. The density matrix $\rho_M$ is then reconstructed via MLE, with state fidelity computed as 
\begin{equation}
F=\left({\rm Tr}\sqrt{\sqrt{\rho_M}\rho_{i}\sqrt{\rho_M}}\right)^2,\label{eq5}
\end{equation}
where $\rho_i$ is idea state.

\subsection*{7. Quantum process tomography}\label{sm_sec7}
Quantum channels provide the natural framework for characterizing high-dimensional quantum storage processes. For a $d$-dimensional input state $\rho$, the storage operation is formally described by a completely positive trace-preserving (CPTP) map $\mathcal{E}$, such that the output state is given by $\mathcal{E}(\rho)$
\begin{equation}
    \mathcal{E}(\rho)=\sum_{m,n=0}^{d^2} \chi_{mn} \hat{\lambda}_m\rho\hat{\lambda}_n ^\dagger,\label{eq6}
\end{equation}
where ${\hat{\lambda}_m}$ constitute the Kraus operator basis. For quantum memory characterization, we set $\hat{\lambda}_0=I_{d\times d}$ and let $\hat{\lambda}_i,~(0<i\leq d^2-1)$ denote the generators of SU(d) group. The process matrix $\chi$ with elements $\chi_{mn}$ provides a complete and unique description of the quantum channel $\mathcal{E}$. The process matrix $\chi$ is reconstructed via quantum process tomography (QPT) as follows: 1. Preparing a complete set of input states $\{\ket{\phi_i},1\leq i\leq d^2\}$.  2. Performing QST on the output states using the measurement basis $\{\ket{\psi_i}, 1\leq i\leq d^2\}$. 3. Reconstructing the matrix $\chi$ from the tomographic data. This protocol requires $d^2$ distinct input configurations and $d^2$ projective measurements per output state \cite{o2004quantum}.

The tomographic data can be used to reconstruct the process matrix. The Hermitian matrix $\tilde{\chi}(t)$ is expressed in the form of Eq. \ref{eq3} with parameters $t=\{t_1,t_2,...,t_{d^2}\}$. This parameterization ensures that $\tilde{\chi}$ describes a trace-preservation process satisfying $\sum_{mn}\tilde{\chi}_{mn}\hat{\lambda}^\dagger_m\hat{\lambda}_n=I$. Using the maximum likelihood method (MLE), we can find the optimal estimate of the process matrix by minimizing the function:
\begin{equation}
\begin{aligned}
    f(t)=&\sum_{i,j=1}^{d^2} \frac{1}{C}\left[c_{ij}-C\sum_{m,n=0}^{d^2-1}\langle \psi_j|\hat{\lambda}_m|\phi_i\rangle \langle \phi_i|\hat{\lambda}_n|\psi_j\rangle \tilde{\chi}_{mn}(t)\right]^2\\
    &+\lambda\left[\sum_{m,n,k=0}^{d^2-1}\tilde{\chi}_{mn}(t) \mathrm{Tr}(\hat{\lambda}_n\hat{\lambda}_k\hat{\lambda}_m)-\delta_{k,0}\right],\label{eq7}
\end{aligned}    
\end{equation}
where $c_{ij}$ denotes the photon counts for measurement basis $\ket{\psi_j}$ with input state $\ket{\phi_i}$, $C$ is total count determined from the data, $\lambda$ is a weighting factor and $\delta$ is the Kronecker delta. The parameter $\lambda$ can be adjusted to ensure that the matrix is arbitrarily close to a completely positive map \cite{o2004quantum}. By minimizing the function $f(t)$ through numerical optimization, we obtain the optimal process matrix $\chi$ and and quantum channel $\mathcal{E}$.

In our experimental configuration for four-dimensional quantum storage, the identical set of input states and measurement bases comprises: $\{\ket{C_n},~(\ket{C_n}+\ket{C_m})/\sqrt{2},~(\ket{C_n}+i\ket{C_m})/\sqrt{2},~5\leq n< m\leq 8\}$. This configuration requires 256 distinct measurement settings for photon counting. The reconstructed four-dimensional storage process matrix is presented in Fig. 3(c) of the main text, achieving a fidelity of $97.9\pm 0.2\%$ to $\hat{\lambda}_0$.

\subsection*{8. Quantum channel capacity}\label{sm_sec8}

The classical and quantum information transmission capabilities of a channel are characterized by its classical capacity $C$ and quantum capacity $Q$, respectively. The classical capacity $C$ of a channel represents the highest possible rate-measured in bits per use-at which classical information can be transmitted with vanishing error probability as the number of channel uses approaches infinity \cite{holevo2002capacity,schumacher1997sending}. A lower bound on the classical capacity can be established through the one-shot accessible information:
\begin{equation}
C_1 = \max_{\{P_x, P_x\}} \max_{\{P_y\}} H(X:Y),
\end{equation}
where the maximum is taken over all possible input ensembles ${p_{x},~\rho_{x}}$ and output measurements ${P_{y}}$, and $H(X:Y)$ denotes the mutual information between the random variables $X$ and $Y$.

Defined as the asymptotic limit of infinitely many channel uses, the quantum capacity Q quantifies the maximum number of qubits that can be reliably transmitted per single use of the channel \cite{lloyd1997capacity,shor2002quantum,devetak2005private}. A lower bound on the quantum capacity is given by the one-shot coherent information:
\begin{equation}
Q_1 = \max_{\rho} I_c(\rho),
\end{equation}
where \( T_c(\rho) := S[\mathcal{E}(\rho)] - S_e(\rho, \mathcal{E}) \) is the coherent information of the state \(\rho\) \cite{schumacher1996quantum}, defined in terms of the von Neumann entropy, \( S(\rho) := -\text{Tr}[\rho \log_2 \rho] \), and of the entropy exchange, \( S_e(\rho, \mathcal{E}) := S[(\mathcal{T}_R \otimes \mathcal{E})(|\Psi_\rho\rangle\langle\Psi_\rho|)] \), with \( |\Psi_\rho\rangle \) being any purification of \(\rho\) using a reference quantum system \( R \).

To demonstrate the advantage of our system over conventional qubit-based memory in terms of storage capacity, we aim to determine the quantum channel capacity of the memory. Using process tomography, we obtained a high-dimensional process matrix capable of simulating arbitrary input and output states. By integrating optimization algorithms, we demonstrate that our 4-path storage scheme achieves a channel capacity of $1.932\pm0.007$, surpassing the theoretical limit of qubit memory (which is 1).

At the same time, the exceptional performance of our quantum memory can be further validated through the entangled Schmidt number. We simulate the storage of a 4-dimensional maximally entangled state and, based on the reconstructed process matrix, calculate an output state fidelity of $F = 0.975$. This high fidelity not only surpasses the threshold for certifying a Schmidt number of 4 ($F > 0.75$) but also robustly demonstrates that our memory achieves a genuine 4-dimensional entanglement capability.

\subsection*{9. The conditional fidelity}\label{sm_sec9}
\begin{figure}[h]
    \centering
    \includegraphics[width=1\linewidth]{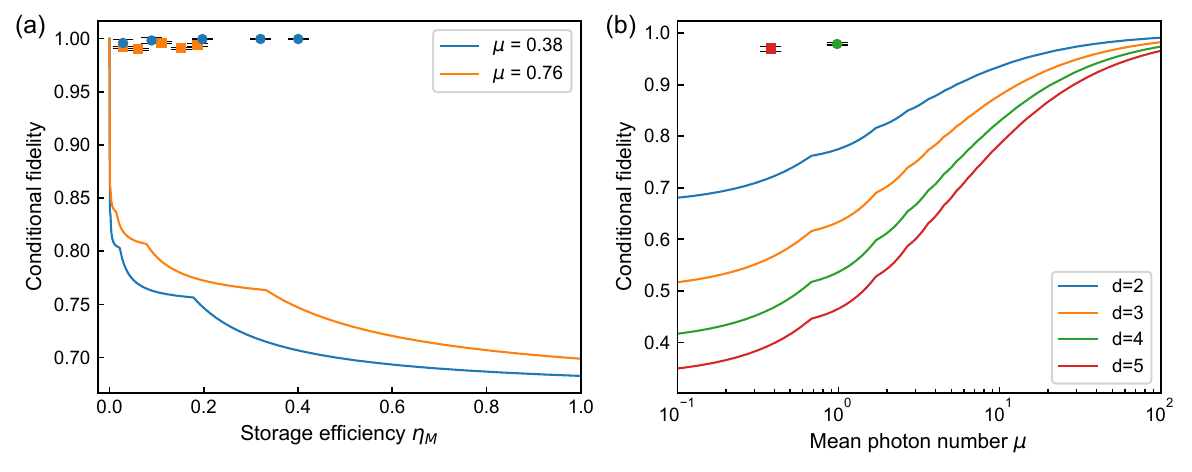}
    \caption{(a) Maximum achievable fidelity for classical storage of a qubit, as a function of memory efficiency $\eta_M$ at mean photon numbers $\mu=0.38$ (blue solid) and $\mu=0.76$ (orange solid). Experimental data: path qubits storage (blue circles) and time-bin qubits storage (orange squares). (b) Maximum achievable fidelity for classical storage of a qudit, as a function of mean photon number $\mu $ with a memory efficiency $\eta_M=0.3$. Experimental data: 4-dimensional quantum storage (green circles) and 5-dimensional quantum storage (red squares).}
    \label{sm_fig7}
\end{figure}

To certify genuine quantum storage, we benchmark the process against the highest fidelity attainable with any classical measure-and-prepare strategy, accounting both for finite storage efficiency and the Poissonian statistics of the weak coherent input. For a system of $N$ photonic qubits, the classical fidelity limit is $F=\frac{N+1}{N+2}$. For a general $d$-dimensional quantum system, the optimal state estimation fidelity is $F_d=\frac{N+1}{N+d}$. When high-dimensional quantum states are encoded via weak coherent states, it should be noted that photon numbers follow a Poisson distribution $P(\mu,N)=\mathrm{e}^{-\mu}\mu^N/N!$ with mean photon number $\mu$. The maximum achievable fidelity is given by
\begin{equation}
F_{\text{class}}^d(\mu)=\sum_{N\geq 1}\frac{N+1}{N+d}\frac{P(\mu,N)}{1-P(\mu,0)},\label{eq8}
\end{equation}
This fidelity threshold holds for quantum memory with unit efficiency. If storage efficiency $\eta_M<1$, the classical fidelity bound for $d$-dimensional systems is given by  
\begin{equation}
F_{\text{class}}^d(\mu,\eta_M)=\frac{\frac{N_{\text{min}}+1}{N_{\text{min}}+d}\gamma+\sum_{N\geq N_{\text{min}}+1}\frac{N+1}{N+d}P(\mu,N)}{\gamma+\sum_{N\geq N_{\text{min}}+1}P(\mu,N)},\label{eq9}
\end{equation}
where $N_{\text{min}}$ is the minimum $i$ that satisfies $\sum_{N\geq i+1}P(\mu,N)\leq [1-P(\mu,0)]\eta_M$, and $\gamma=(1-P(\mu,0))\eta_M-\sum_{n\geq N_{\text{min}+1}}P(\mu,n)$ \cite{gundougan2015solid,dong2023highly}. Fig. \ref{sm_fig7}(a) plots the classical fidelity bound as a function of storage efficiency for mean photon numbers $\mu=0.38$~ and $\mu=0.76$~, corresponding to the inputs used for path and time-bin qubits, respectively. The experimentally measured fidelities (symbols) lie well above these bounds, confirming genuine quantum storage for both encodings. Fig. \ref{sm_fig7}(b) shows the classical fidelity limit for dimensions d = 2–5 at the measured efficiency $\eta_M=0.3$ (1 \us~high-dimensional storage). Our high-dimensional memories deliver fidelities that surpass the classical bound by substantial margins across all tested dimensions.

\end{document}